\documentclass[useAMS,usenatbib]{mn2e}

\usepackage{graphicx}
\usepackage{epsfig} 
\usepackage{amssymb}
\usepackage{textcomp}
\usepackage{bm}

\title{Knock-on processes in superfluid vortex avalanches and pulsar glitch statistics}
\author[L. Warszawski and A. Melatos]{L. Warszawski$^{1}$\thanks{lila@unimelb.edu.au (L
W)} and A. Melatos$^{1}$\thanks{amelatos@unimelb.edu.au (AM)}\\
$^{1}$School of Physics, University of Melbourne, Parkville, VIC 3010, Australia\\}
\begin{document}

\date{\today}
\maketitle

\begin{abstract}
A framework is presented for a statistical theory of neutron star glitches, motivated by the results emerging from recent
Gross-Pitaevskii simulations of pinned, decelerating quantum condensates. It is shown that the observed glitch size distributions cannot be
reproduced if superfluid vortices unpin independently via a Poisson process; the central limit theorem yields a narrow Gaussian for the size distribution,
instead of the broad, power-law tail observed. This conclusion is not altered fundamentally when a range of pinning potentials is included, which leads to excavation of the
potential distribution of occupied sites, vortex accumulation at strong pinning sites, and hence the occasional, abnormally large glitch.  Knock-on processes are therefore needed to make the unpinning rate of a vortex conditional on the pinning state of its near and/or remote neighbours, so that the Gaussian size distributions resulting generically from the central limit theorem are avoided. At least two knock-on
processes, nearest-neighbour proximity knock-on and remote acoustic knock-on, are clearly evident in the Gross-Pitaevskii
simulation output. It is shown that scale-invariant (i.e. power-law) vortex avalanches occur when knock-on is included, provided that two specific
relations hold between the temperature and spin-down torque. This fine tuning is unlikely in an astronomical setting, leaving the overall problem partly unsolved.  A state-dependent Poisson formalism is presented which will form the basis of future studies in this area.
\end{abstract}

\section{Introduction}

Pulsar glitches --- the discrete, randomly-timed jumps in the spin frequency of a pulsar --- are characterised by power-law size and exponential waiting-time distributions \citep{Wong:2001p192,Melatos:2008p204}.  These statistics point to an underlying collective process, similar to other complex systems such as neural and social networks \citep{Worrell:2002}, soil moisture balance \citep{Porporato:2004p8491,Daly:2007p8447}, electricity grids \citep{Carreras:2002p8220}, forest fires \citep{Turcotte:1999}, earthquakes \citep{Drossel:1996} and many more \citep{Goss:1989,Schonfisch:1999,Cornforth:2005,Boerlijst:1991,Suki:1994,Wu:2010}.  All of these systems respond to a global driver via local, stress-releasing interactions between discrete elements and may be modelled using many-body techniques.  

Pulsar glitches are caused by the unpinning and near-simultaneous outward motion of many (up to $\sim 10^{14}$) quantised superfluid vortices in the pulsar inner crust \citep{Anderson:1975p84}.  Individual, non-interacting vortices unpin stochastically according to a Poisson process, whose rate depends on the system temperature, local pinning forces \citep{Hanggi:1990,Hakonen:1998p10787}, local vortex density, and the relative velocity of the superfluid and pulsar crust.  In addition to these factors, collective unpinning events are catalysed by unpinned vortices, which raise the unpinning rate of other vortices (nearby or distant, depending on the knock-on mechanism).  These collective, avalanche-like events, in which vortices execute jerky, stick-slip motion, proceed in a similar way to cascading failures in electrical grids and other large, networked systems \citep{Carreras:2009p8216}, via an underlying threshold-driven branching process.  They are characterised by a critical state in which the event size distribution is a power law with exponent $-3/2$.  This simple power-law characterisation is most appropriate to a system in which the event rate is constant.  For systems like pulsars in which the event rate depends on some constantly evolving system parameter \citep{Daly:2007p8447}, a more general formulation is required.

The occurrence of a glitch brings the superfluid and crust closer to corotation and hence reduces the global unpinning rate.  This feedback regulates the state-dependent vortex unpinning rate, which, when combined with the persistent electromagnetic spin down of the superfluid crust, leads to a self-organising system, whose evolution resembles that of a slowly driven sandpile or a forest fire \citep[e.g.][]{Drossel:1996}.  That is, vortex unpinning decelerates the superfluid, reducing the crust-superfluid lag and hence temporarily decreasing the unpinning rate, which in turn increases over the long term in response to the electromagnetic spin-down torque.  

This paper asks whether a system of pinned vortices that unpin according to a variable-rate Poisson process, whose evolution is governed by pulsar-like rules, exhibits unpinning avalanches like those observed.  We build on the coherent noise model proposed by \cite{Melatos:2009p4511}, which successfully generates power-law-distributed glitches, with the aim of deducing the waiting-time distribution endogenously, instead of putting it in by hand.  Vortex knock-on, as observed in quantum mechanical simulations (see \cite{Warszawski:2010individual}), provides a natural catalyst for unpinning avalanches.  We propose a statistical model that self-consistently describes feedback between the superfluid and pulsar crust and introduces three novel features that reflect the latest understanding of pulsar interiors:  stochastic unpinning, knock-on processes, and multiple pinning energies.  The consequences of each feature are is assessed independently. We compare our analytic results to the output of Monte-Carlo simulations to test the underlying assumptions and explore regimes that cannot be treated analytically. 

The paper is organised as follows.  In Sec.~\ref{sec:glitchmodel} we summarise the (un)pinning model of pulsar glitches and the challenges and paradoxes that it faces.  Section~\ref{sec:unpinningknock} formulates vortex unpinning as a Poisson process.  The feedback mechanism that relates vortex unpinning to changes in crust rotation is described in Sec.~\ref{sec:feedback}.  In Sec.~\ref{sec:MCsimple} we summarise results from an asynchronous Monte-Carlo automaton, where the variable time step is governed by the shortest waiting time between vortex unpinning events, that realises the processes described in Sec.~\ref{sec:glitchmodel}--\ref{sec:feedback}.  We investigate the effect of pinning at one unique energy versus multiple energies in Sec.~\ref{subsec:homoopin} and \ref{sec:heteropinning} respectively.  In Sec.~\ref{sec:branching} we outline possible ways that unpinned vortices can catalyse unpinning avalanches, giving rise to a branching process, and describe how branching can be incorporated into both the analytic and Monte-Carlo treatments.  It has proved useful to develop the Monte-Carlo and analytic descriptions side by side.  In Sec.~\ref{sec:master} we describe how a master equation can be used to model the probability density function of the state of the system.  The problem is formulated in terms of a state-dependent process with Poisson transition statistics \citep{Daly:2007p8447,Wheatland:2008p8329}.  Conclusions and prospects for future work are contained in Sec.~\ref{subsec:summaryknock}.

\section{Vortex unpinning paradigm}\label{sec:glitchmodel}

\subsection{Standard model}\label{subsec:collective}

A pulsar's inner crust contains a neutron superfluid flowing through a crystalline lattice of nuclei. The superfluid rotates via the formation of many ($\sim 10^{19}$) quantised vortices, whose unperturbed configuration is an equally spaced, hexagonal Abrikosov lattice \citep{Donnelly}. Each vortex carries a quantum of circulation and generates a solenoidal velocity field.  Macroscopically, the velocity fields from many vortices add vectorially to mimic rigid body rotation.  It is energetically favourable for a vortex core to sit atop a column of impurities (such as nuclear lattice sites \citep{Jones:1991p3949}), such that the volume from which superfluid is excluded is minimised.  The strength of this phenomenon, known as pinning, is density dependent \citep{Pizzochero:1997p45,Donati:2006p32,Pizzochero:2011}.  In addition, the extent of vortex pinning depends on the relative strength of the vortex-nucleus interaction and vortex tension \citep{Link:1993p47,Haskell:2011}.

As the pulsar spins down electromagnetically, pinning locks in the vortex lattice configuration, preventing the superfluid from decelerating commensurately with the crust.  Differential rotation between the ambient superfluid and the pinned vortices, which necessarily corotate with the crust, exerts a transverse Magnus force on the vortex that, when sufficiently strong, unpins the vortex.  The unpinning event and subsequent outward vortex motion release angular momentum to the crust, causing a glitch \citep{Anderson:1975p84,Alpar:1981p18}. 
 
The simplest form of the model outlined above is deterministic.  That is, a vortex unpins when the shear between the container and superfluid exceeds a critical threshold and stays pinned when the shear is subcritical.  The Magnus force on pinned vortices grows linearly over time, and all vortices pinned at the same cylindrical radius experience the same force.  Periodic glitches of equal size are direct corollaries of this picture.  However, periodic glitches of equal size are not seen in the data, except in two objects  \citep{Melatos:2008p204,Espinoza:2011}.  The standard picture can be modified to include strong and weak pinning zones \citep{CHENG:1988p180} without treating local fluctuations in pinning conditions at each vortex.  However, whether or not a two-zone mean-field model is sufficient to reproduce the observed glitch size and waiting time distributions is an unanswered question.

\subsection{New physical ideas}  \label{subsec:newphysics}

New physics needs to be added to the vortex unpinning paradigm in order to understand: (1) why some pulsars glitch and others do not; (2) why glitch sizes span up to four decades in an individual pulsar; (3) why there are large glitches involving the simultaneous unpinning of $\sim 10^{12}$ vortices, rather than a steady \emph{trickle} of unpinning events; (4) why vortices skip over $\sim 10^9$ pinning sites before repinning; (5) if (how) large-scale inhomogeneities develop in a vortex lattice pinned to a homogeneous pinning array; and (6) how waiting times between glitches can be deduced endogenously from a model based on individual vortex unpinning.

Our model addresses many of these questions (repinning is not specifically modelled\footnote{How vortices repin remains an unsolved problem. One important question is how many pinning sites a vortex "skips over" before it repins. Upon examining many movies from Gross-Pitaevskii simulations of a pinned, decelerating condensate, whose results are reported in detail in \cite{Warszawski:2010pulsar}, we tend to observe that vortices skip over
as many pinning sites as necessary to maintain roughly one Feynmann distance between each other. However, there are considerable fluctuations about this mean behaviour.  Sometimes, a group of vortices unpin and leaves behind a gap in the Abrikosov array which is metastable and takes many avalanche "cycles" to fill. Other times, an unpinned vortex travels further than expected after grazing several pinning sites like a golf ball rimming the cup \citep{Warszawski:2010individual}. None of these special situations are
captured in the present paper and should be included in a more complete model.}) by incorporating the following three novel features:  (1) individual vortices unpin explicitly according to a Poisson process whose rate is governed by the global superfluid-crust shear; (2) there is a range of available pinning energies, representing the myriad imperfections in the pulsar crustal lattice (we find that deep pinning wells are systematically overpopulated compared to shallow wells); and (3) unpinned vortices catalyse knock-on unpinnings at neighbouring pinning sites, which we model as a branching process.

Domino-like knock-on unpinning, leading to collective vortex unpinning, is capable of explaining the observed scale-invariant dynamics of pulsar glitches.  Numerical solutions of the Gross-Pitaevskii equation show that knock-on events arise naturally in a pinned, rotating condensate \citep{Warszawski:2010individual}, and the resulting glitches span several decades in glitch size \citep{Warszawski:2010pulsar}. Knock-on events are triggered by vortex-vortex proximity, or acoustic radiation emitted during a nearby or remote repinning event.  Knock-on stands in contrast to individual unpinnings characteristic of vortex creep \citep{Alpar:1984p6781}, although both types of events can occur simultaneously in the same system.

The dynamics of individual vortices traversing an inhomogeneous pinning landscape is an active field of research.  Gross-Pitaevskii simulations show that vortices do not necessarily repin at the next available pinning site [see \cite{Warszawski:2010individual} and \cite{Warszawski:2010pulsar}].  The exact trajectory depends on the superfluid-crust lag and the strength of pinning.  \cite{Jones:1991p92}, \cite{Jones:1998p34} and \cite{Link:2009p9063} showed that there is a critical velocity below which a vortex travelling past a pinning site is immobilised. Its existence implies that the crust-superfluid lag necessary to overcome the pinning force must be large enough to propel a vortex past many pinning sites.  Furthermore, in order to catalyse glitches of the size observed, vortices that unpin must move a distance comparable to the mean inter-vortex spacing ($\sim 1\,\rm{mm}$) before repinning, which again involves bypassing many available pinning sites. 

Previous models, such as the avalanche model in \cite{Warszawski:2008p4510} and the coherent noise model in \cite{Melatos:2009p4511}, suffer from two major drawbacks.  The avalanche model requires inhomogeneities in the pinned vortex distribution on scales many orders of magnitude greater than the average intervortex separation, contravening nuclear structure calculations \citep{Donati:2006p32,Avogadro:2007p51}.  The coherent noise model, which is spatially uniform by construction, offers an elegant way around this problem, by allowing pinning strengths to vary from site to site.  However, in the form presented in \cite{Melatos:2009p4511}, the waiting-time distribution is entered by hand, by fitting to observational data; it is not derived self-consistently.

\section{Individual vortex unpinning as a Poisson process}\label{sec:unpinningknock}
Pinned vortices move relative to the bulk superfluid, experiencing a Magnus force per unit length equal to
\begin{equation}
\label{eq:Magnus}
 \mathbf{F}_{\rm{M}}=-\rho\bm{\kappa}\times\left(\mathbf{v_{\rm{L}}}-\mathbf{v_{\rm{s}}}\right)~,
\end{equation}
where $\bm{\kappa}=(h/m)\hat{\bm{\Omega}}_{\rm{s}}$ is the quantum of circulation directed along the rotation axis $\hat{\bm{\Omega}}_{\rm{s}}$, $\rho$ is the superfluid density, and $\mathbf{v_{\rm{L}}}-\mathbf{v_{\rm{s}}}$ is the velocity difference between the vortex line (which corotates with the crust) and the ambient superfluid.  In the many-vortex limit, the superfluid mimics rigid-body rotation with angular velocity $\Omega_{\rm{s}}$.  Assuming that the vortices are straight and parallel and intersect the equatorial plane of the pulsar \footnote{While it is traditional in neutron star literature to consider rectilinear vortices, and we continue that tradition here, simulations show that at characteristic pulsar rotation rates, the flow in the inner crust is turbulent, and hence a tangle of vortices is likely \citep{Tsubota:2000,Peralta05,Peralta06a,Melatos:2007p158,Andersson:2007p196,Glampedakis:2008p7,Melatos:2010turb}. }, we can write the Magnus force at radius $R$ as
\begin{equation}
\label{eq:Magnus_omega}
  F_{\rm{M}}=\rho \kappa R\Delta\Omega_{\rm{cr}}\frac{\Delta\Omega}{\Delta\Omega_{\rm{cr}}}~,
\end{equation}
with $\Delta\Omega = \Omega_{\rm{s}}-\Omega_{\rm{c}}=|\mathbf{v_{\rm{L}}}-\mathbf{v_{\rm{s}}}|/R$, where $\Omega_{\rm{c}}$ is the angular velocity of the crust, and $\Delta\Omega_{\rm{cr}}$ is the maximum differential angular velocity that a vortex pinned with energy $E_{\rm{p}}$ can withstand before unpinning \citep{Alpar:1984p6781}.   $\Delta\Omega_{\rm{cr}}$ is a function of $E_{\rm{p}}$, the superfluid coherence length, $\xi$, and the pinning site separation $b$, according to
\begin{eqnarray}
 \Delta\Omega_{\rm{cr}}&=&\frac{E_{\rm{p}}}{b\xi\rho\kappa R} \\
 &=& 10^{-3}\left(\frac{E_{\rm{p}}}{1\,\rm{MeV}}\right) \left(\frac{100\,\rm{fm}}{b}\right)\left(\frac{10\,\rm{fm}}{\xi}\right)\left(\frac{10^{17}\,\rm{kg\,m}^{-3}}{\rho}\right)\left(\frac{10^4\,\rm{m}}{R}\right)\,\rm{rad\,s}^{-1}~.
\end{eqnarray}
Following \cite{Link:1993p47}, we use Eq.~(\ref{eq:Magnus_omega}) to write the reduced pinning force, $F^{\prime}$, on a pinned vortex that rotates differentially with the superfluid, as 
\begin{eqnarray}
\label{eq:FpFm}
 F^{\prime}=F_{\rm{p}}-F_{\rm{M}}
			&=&\rho\kappa R\Delta\Omega_{\rm{cr}}\left(1-\frac{\Delta\Omega}{\Delta\Omega_{\rm{cr}}}\right)~.
\end{eqnarray}

It is useful to describe vortex unpinning by analogy with Kramers escape problem of a classical or quantum mechanical particle confined in a potential well \citep{Gardiner:2002}.  That is, the Boltzmann escape or tunnelling probability increases with increasing system temperature and any local and global energy biases, including the differential angular velocity entering Eq.~(\ref{eq:FpFm}).  Hence, we employ the Arrhenius formula to write down the unpinning rate, $\theta$, of an individual vortex in a system with reciprocal temperature $\beta=(k_{\rm{B}}T)^{-1}$, pinned with energy $E_{\rm{p}}$, as \citep{Hanggi:1990,Chevalier:1993p7949} 
\begin{equation}
\label{eq:rate}
 \theta(E_{\rm{p}},\gamma)=\Gamma_0 e^{-\beta E_{\rm{p}}\gamma}\,\rm{s}^{-1}~,
\end{equation}
where $\gamma = 1-\Delta\Omega/\Delta\Omega_{\rm{cr}}$ parametrises the reduction in pinning energy due to differential rotation, and $\langle\tau\rangle=[\theta(E_{\rm{p}},\gamma)]^{-1}$ is the mean waiting time between unpinning events for a single vortex.  The unpinning rate for a system of $N_{\rm{v}}$ identical vortices is $N_{\rm{v}}\theta(E_{\rm{p}},\gamma)$.

The waiting times between unpinning events, $\tau$, are distributed according to an exponential probability density function (PDF),
\begin{equation}
\label{eq:poftauconst}
p(\tau;E_{\rm{p}},\gamma) = \theta(E_{\rm{p}},\gamma)\exp\left[-\theta(E_{\rm{p}},\gamma)\tau\right]~,
\end{equation}
for a constant-rate (fixed $\gamma$) process.  For a variable-rate process, $\theta$ is a function of time through $\gamma(t)$, and hence Eq.~(\ref{eq:poftauconst}) becomes \citep{Wheatland:2008p8329}:
\begin{equation}
\label{eq:poftau}
 p(\tau;E_{\rm{p}},\gamma,t)=\theta[E_{\rm{p}},\gamma(t+\tau)] \exp\left\lbrace-\int_{t}^{t+\tau}dt^{\prime}\,\theta[E_{\rm{p}},\gamma(t+t^{\prime})]\right\rbrace~.
\end{equation}
The integral in the exponent in Eq.~(\ref{eq:poftau}) accounts for the increasing likelihood that unpinning occurs as time passes due to the accumulating crust-superfluid lag.  

In the simple case of uncorrelated unpinning events, the PDF of event sizes is trivially
\begin{equation}
p(n) = \delta(n-1)~,
\label{eq:pdNsingle}
\end{equation}
where $n$ is the number of vortices that unpin.  Equation~(\ref{eq:pdNsingle}) highlights the fact that there are no avalanches.  Individual unpinning events can still cluster in time, as in any Poisson process.  Unlike knock-on-mediated avalanches (see Sec.~\ref{sec:branching}), however, where even a system-spanning avalanche lasts much less than the typical time between events, and it is obvious what constitutes a single avalanche, there is no preferred time-scale for identifying complete events (except the trivial one of recording unpinnings one by one, as above).  In a pulsar, many individual unpinnings occur in a typical observational time window $\Delta t$.  Binning events according to the observational resolution $\Delta t$, we obtain the familiar Poisson result
\begin{equation}
p(n) = \exp(-\theta\Delta t)\frac{\left(\theta\Delta t\right)^n}{n!}~,
\end{equation}
which tends to a Gaussian distribution for large $\Delta t$, and $\theta$ remains the unpinning rate for a single vortex.  The Gaussian conflicts with data in all pulsars except Vela and PSR J0537-6910 \citep{Melatos:2008p204}.

\begin{figure}
\includegraphics[scale=0.36,angle=90]{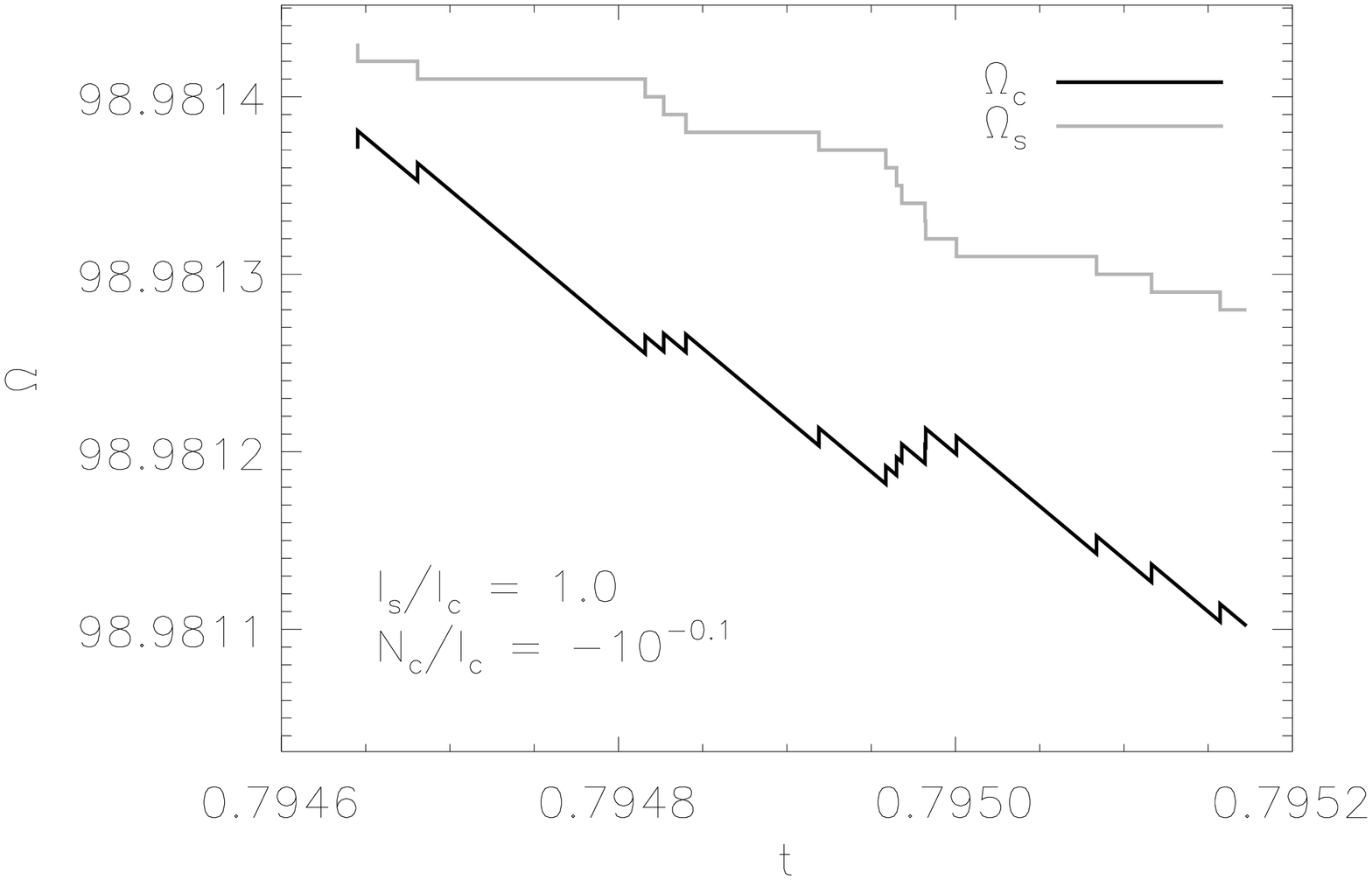}
\includegraphics[scale=0.36,angle=90]{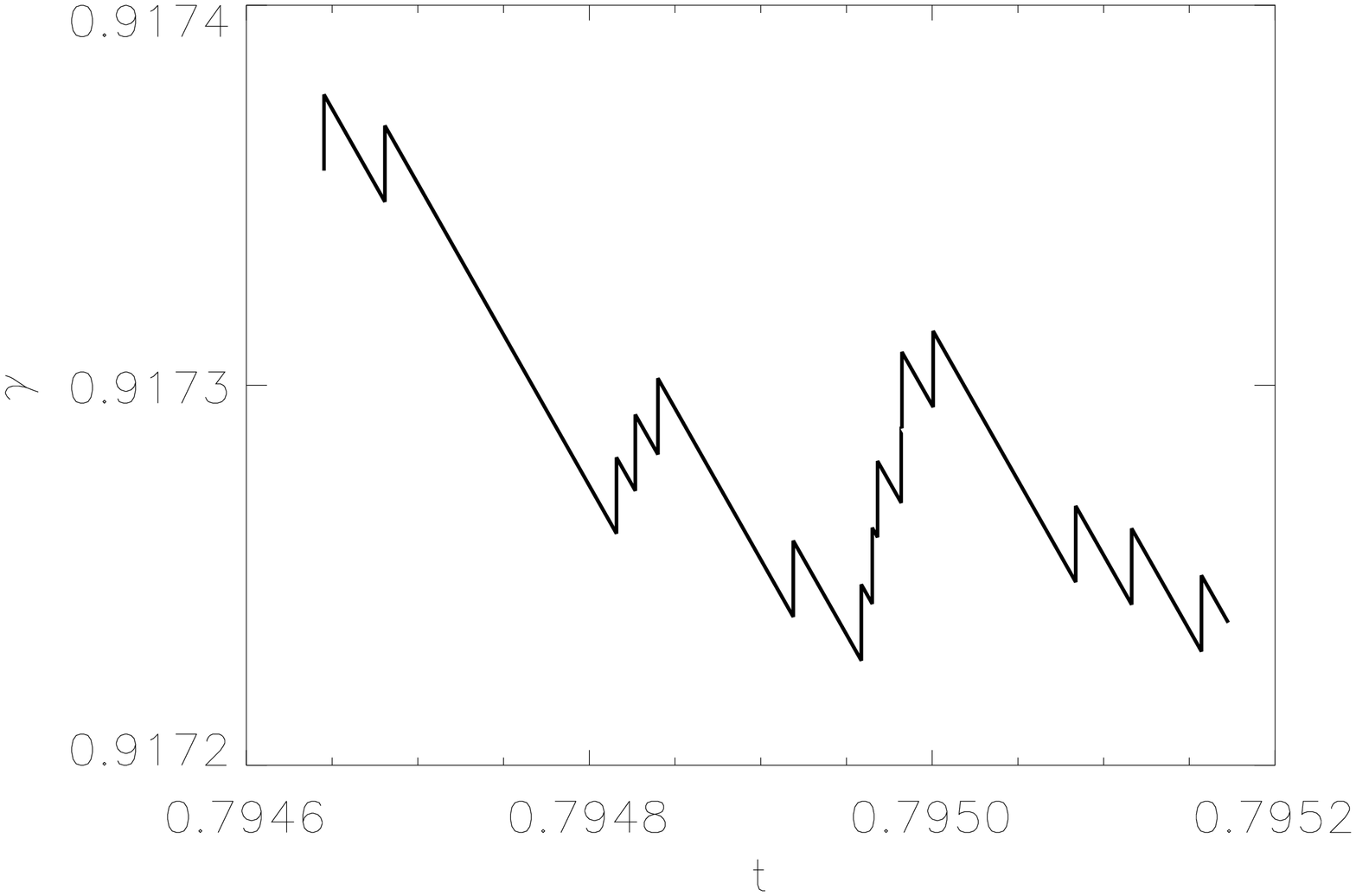}
\caption{Results from a mock simulation, in which power-law distributed glitch sizes and exponential waiting times are superposed on a uniformly decelerating crust.  Equation~(\ref{eq:feedback}) describes how changes in $\Omega_{\rm{c}}$ and $\Omega_{\rm{s}}$ are related.  \emph{Left}: Angular velocity as a function of time $\Omega_{\rm{c}} (t)$ for the crust ($\Omega_{\rm{c}}$, \emph{black} curve) and superfluid ($\Omega_{\rm{s}}$, \emph{grey} curve; arbitrarily shifted down by $0.706$ for comparison).  The superfluid spins down in a step-wise fashion, corresponding to vortex unpinning, whilst the crust angular velocity decreases linearly with time, accelerating instantaneously when the superfluid decelerates. \emph{Right}: Differential rotation parameter as a function of time, $\gamma(t)$, for the same interval graphed in the \emph{left} panel.  $\gamma(t)$ decreases linearly between unpinning events, and increases in discrete jumps when vortices unpin.  Simulation parameters:  $I_{\rm{s}}=I_{\rm{c}}$.}
\label{fig:schematic}
\end{figure}

\begin{figure*}
\includegraphics[scale=0.36,angle=90]{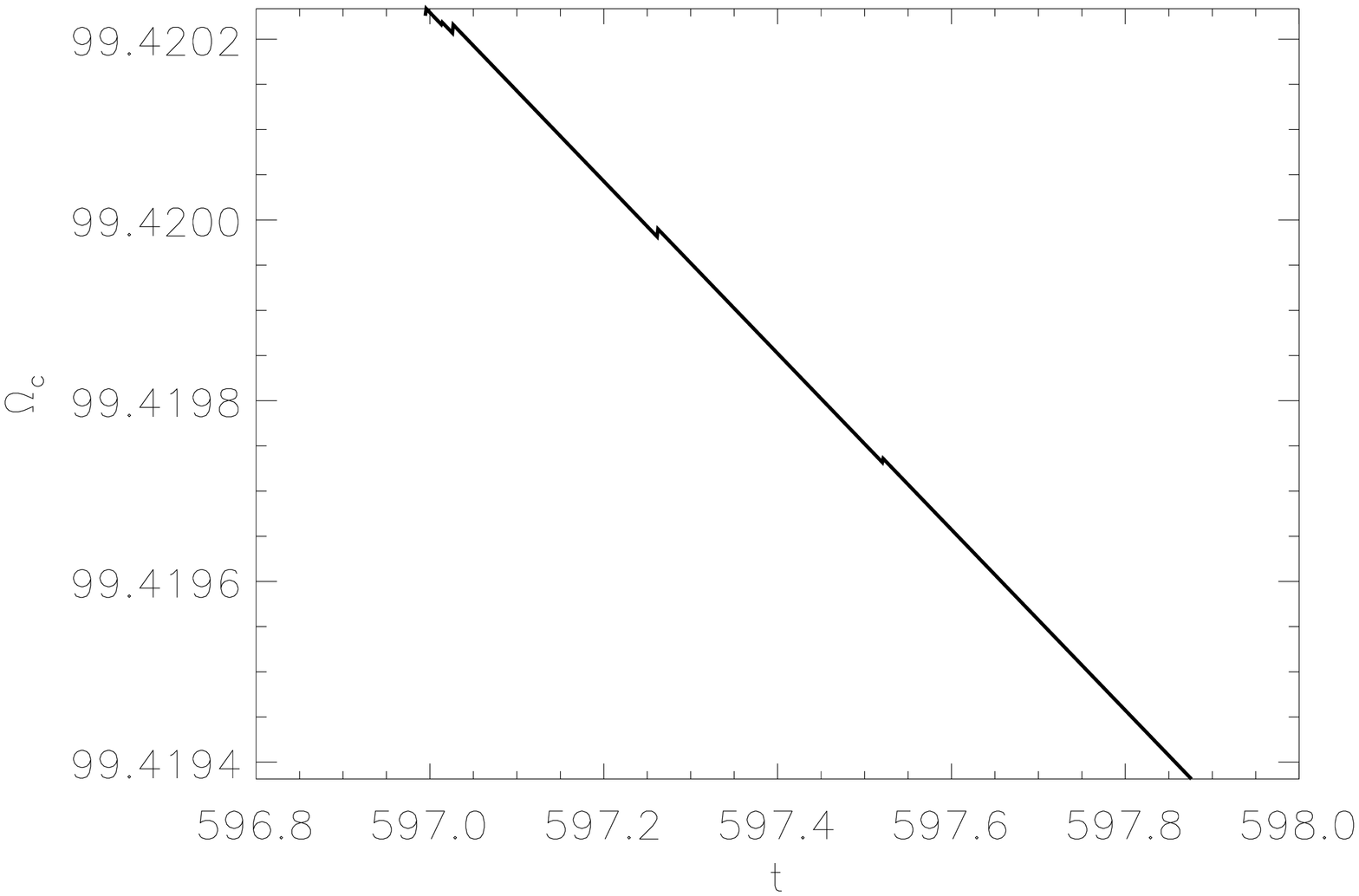}
\includegraphics[scale=0.36,angle=90]{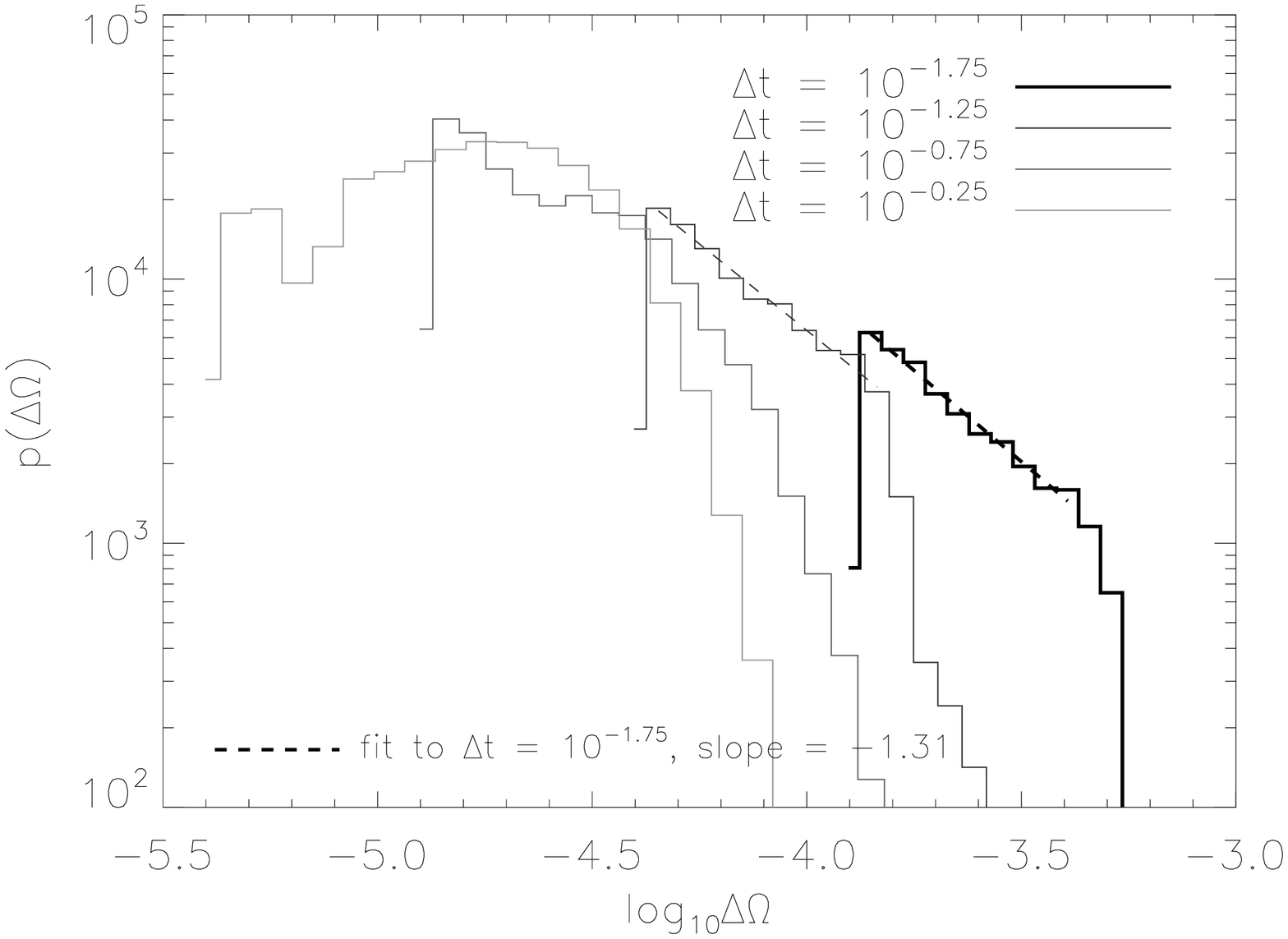}
\caption{Results of a mock simulation showing how the PDF of observed glitch sizes depends on the observation time window for binning, $\Delta t$.  In the underlying simulation, a power-law distribution of glitches (index $-3/2$), separated by exponentially distributed intervals, is superposed on a background of constant deceleration.  \emph{Left}:  Angular velocity of crust as a function of time $\Omega_{\rm{c}}(t)$.  \emph{Right}:  PDFs of `observed' glitch sizes for four time windows, $\Delta t = 10^{-1.75}$, $10^{-1.25}$, $10^{-1.00}$ and $10^{-0.75}$ (\emph{black} to \emph{light grey} curves respectively).  The glitch-finding algorithm is described in Sec.~\ref{subsec:glitchfind}.}
\label{fig:mock}
\end{figure*}

\section{Feedback and self-regulation}\label{sec:feedback}

\begin{figure}
\includegraphics[scale=0.36,angle=90]{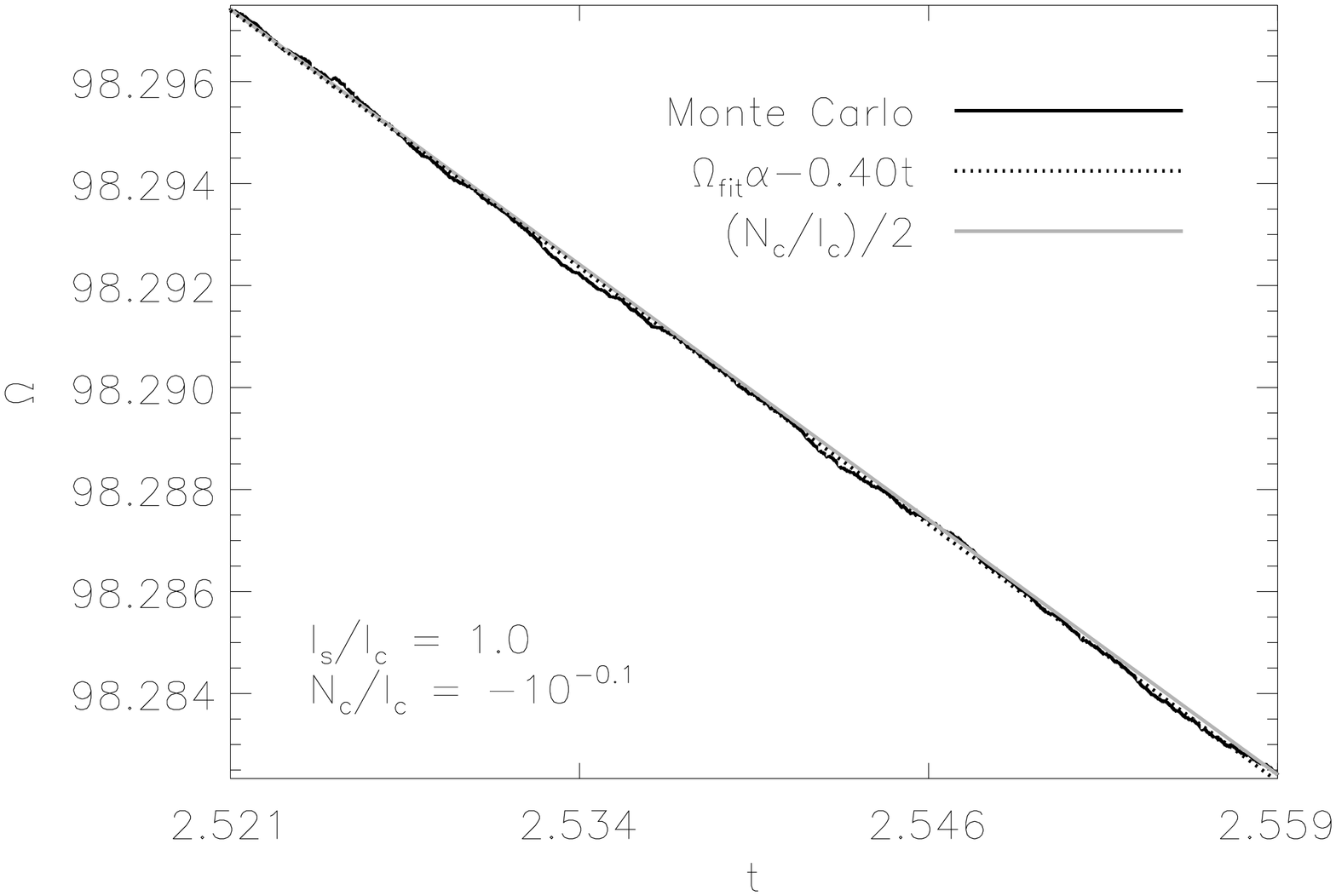}
\includegraphics[scale=0.36,angle=90]{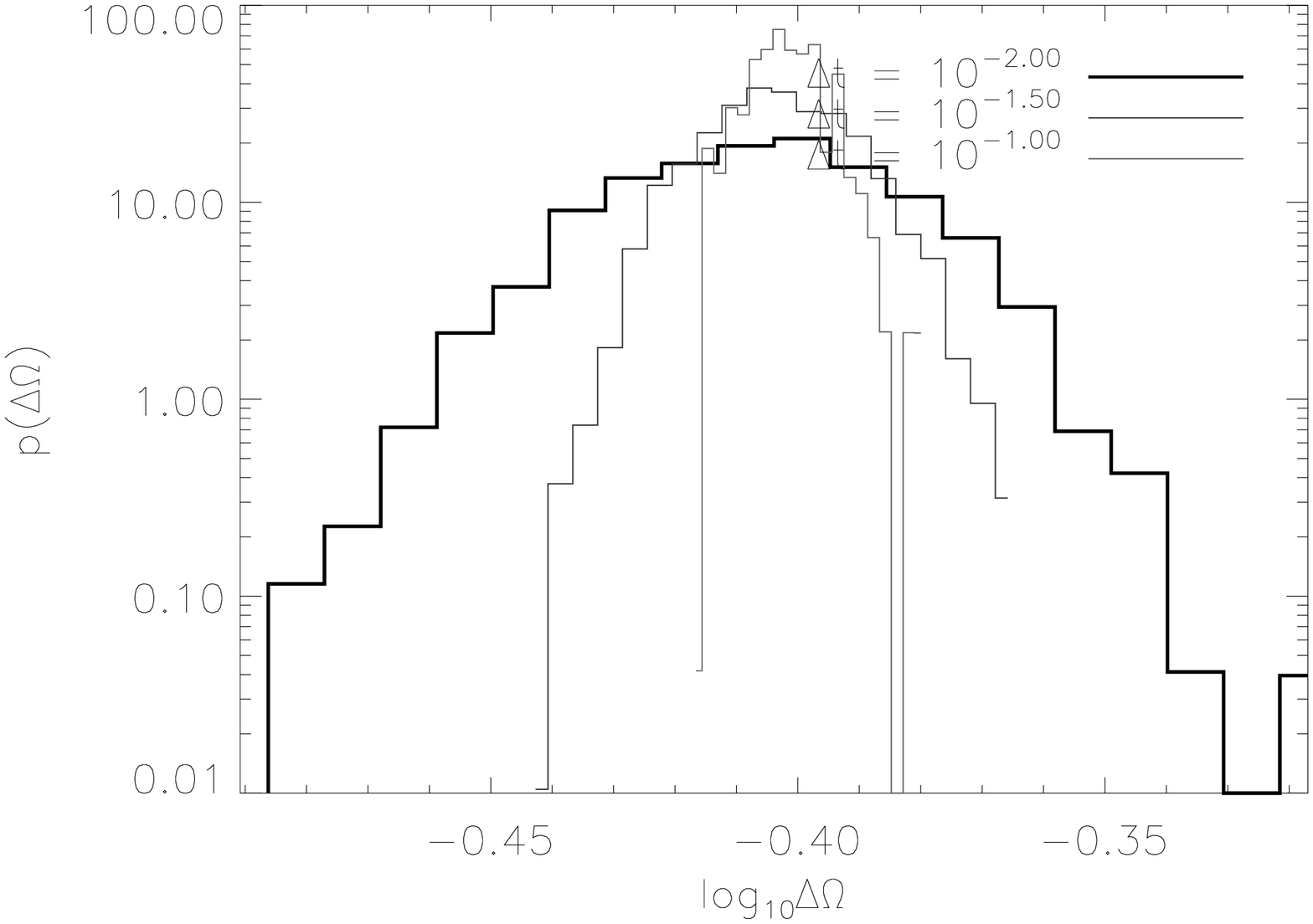}
\includegraphics[scale=0.36,angle=90]{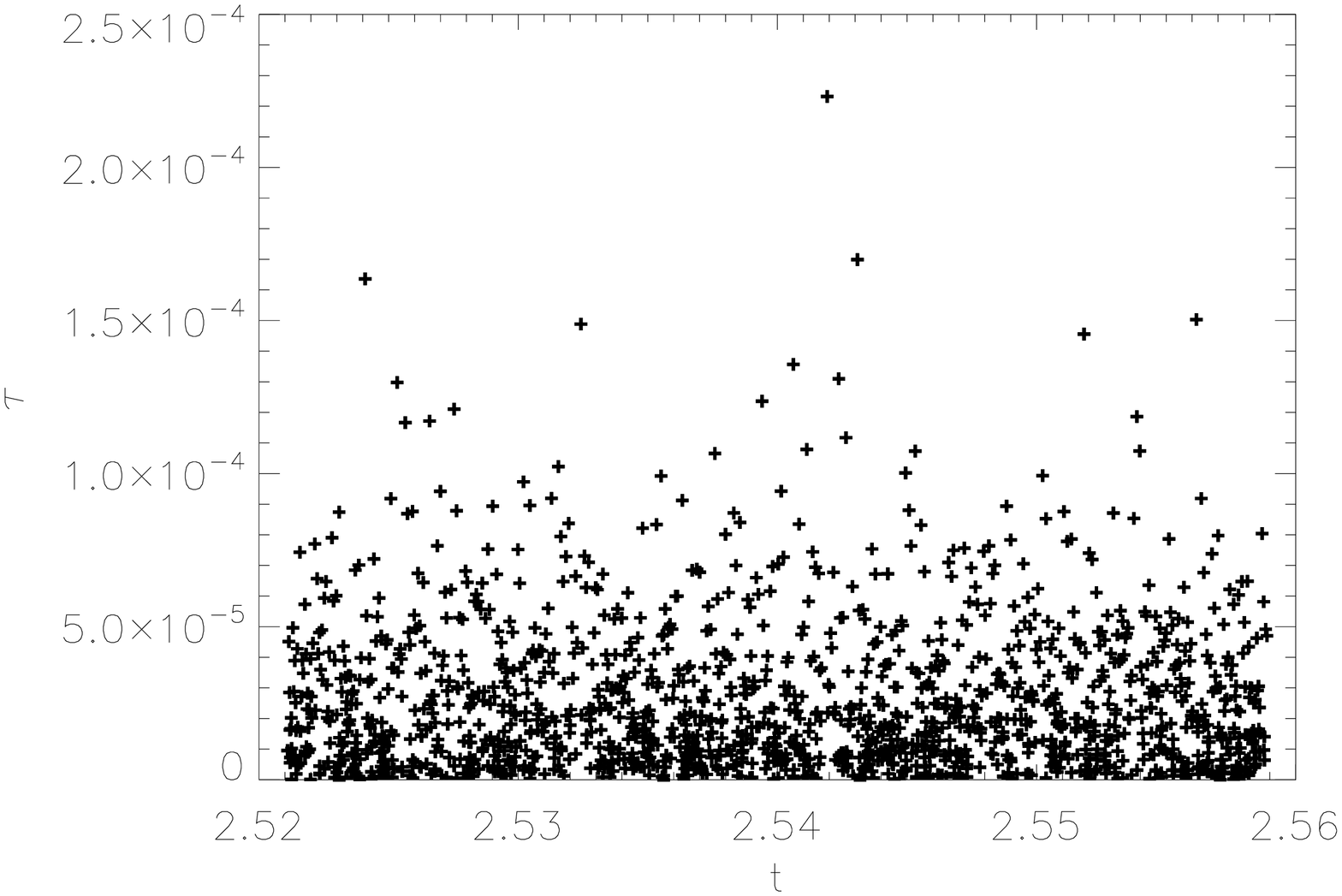}
\includegraphics[scale=0.36,angle=90]{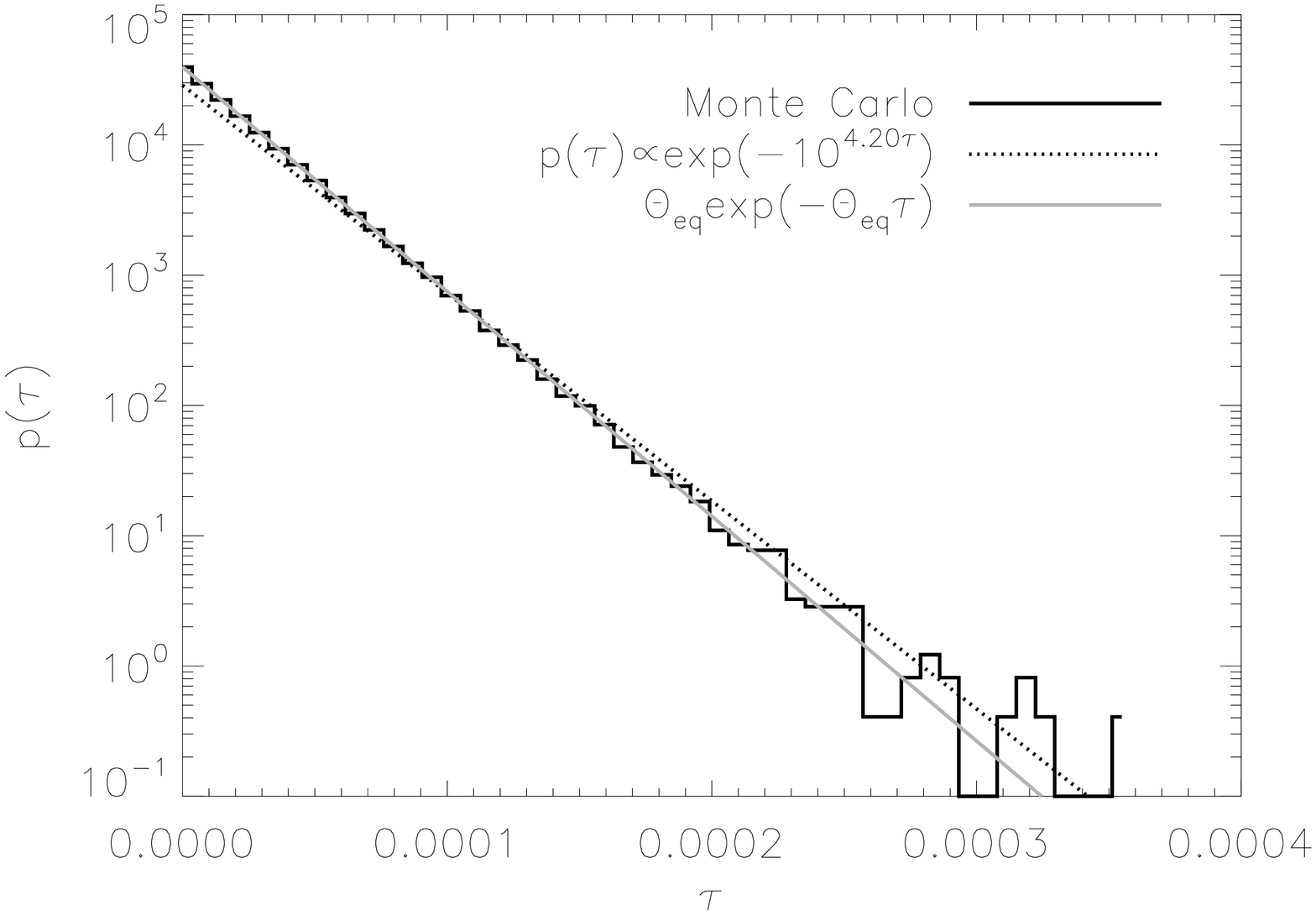}
\includegraphics[scale=0.36,angle=90]{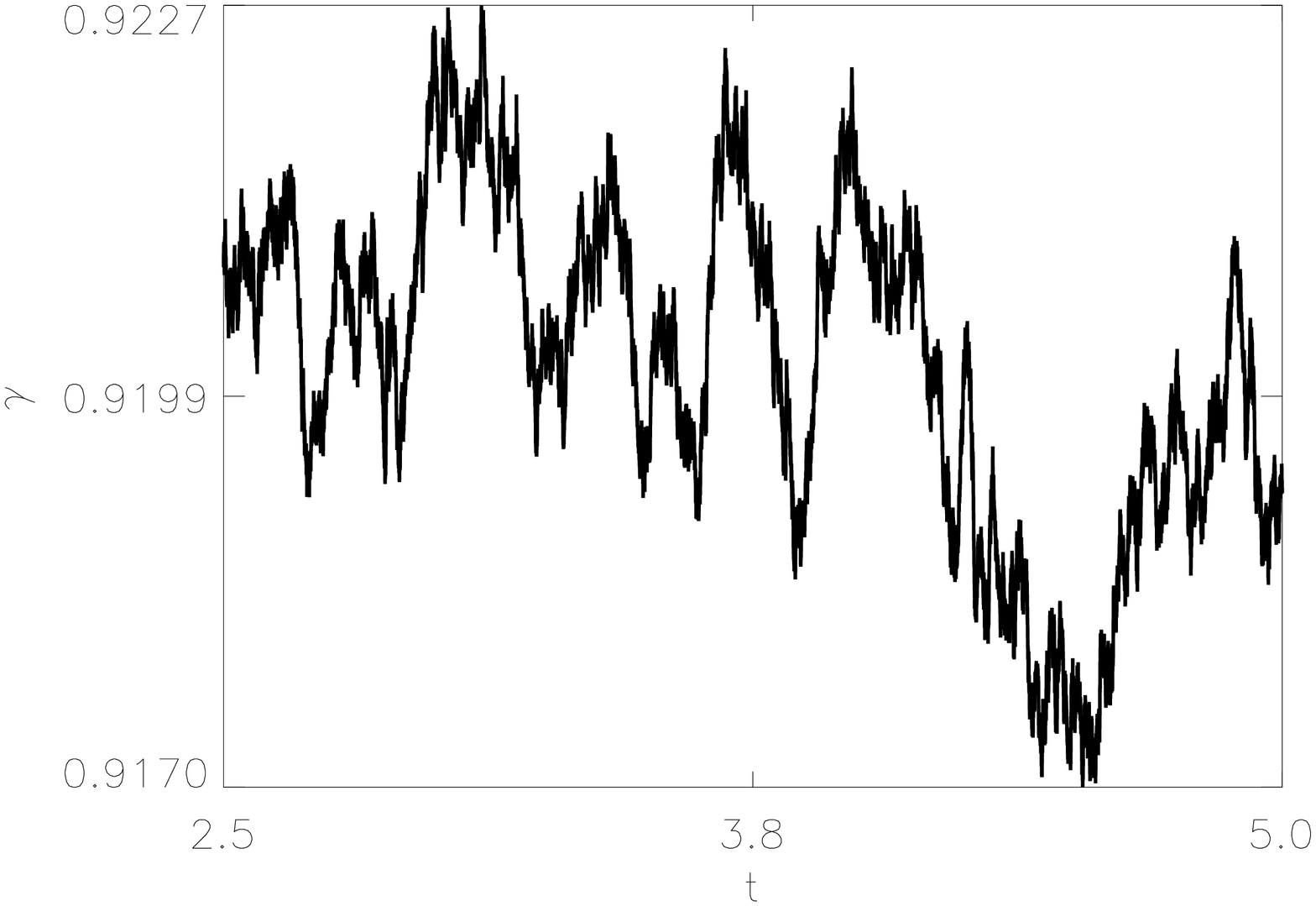}
\includegraphics[scale=0.36,angle=90]{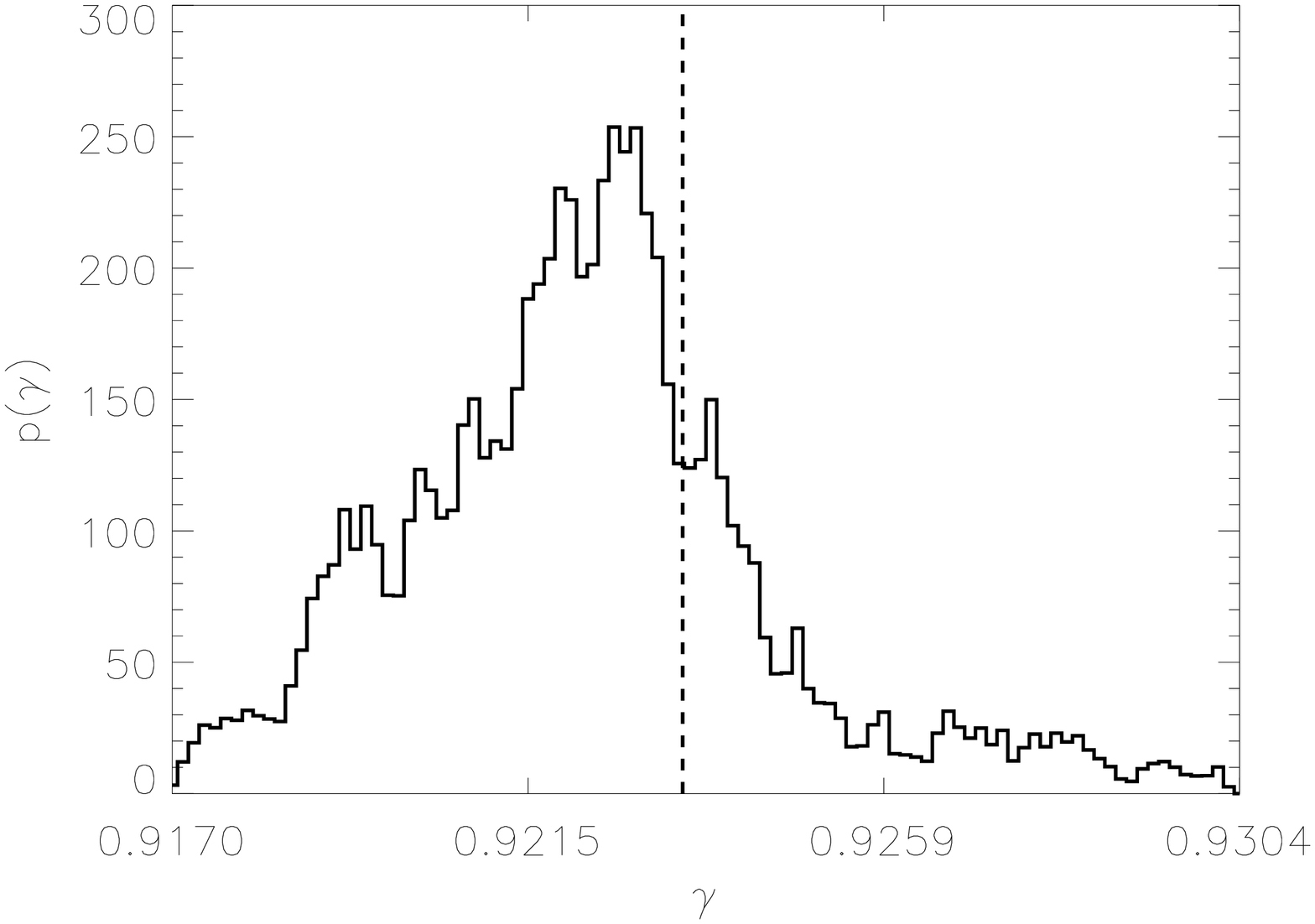}
\caption{Monte-Carlo simulations of a vortex system with a \emph{unique} pinning energy, torque feedback described by Eq.~(\ref{eq:feedback}), and no knock-on avalanches.  The automaton uses the rules described in Sec.~\ref{sec:MCsimple}, for a system of $N_{\rm{v}}=10^3$ vortices with $E_{\rm{p}}=E_0=1.0$, $N_{\rm{c}}/I_{\rm{c}}=10^{-0.1}$, $\alpha=10^{-4}$, $I_{\rm{s}}/I_{\rm{c}}=1.0$ and $N_t=10^6$.  Results are reported as time series and PDFs.  The PDFs are constructed by sampling the time-varying quantity, and making a histogram of the sampled values.   The glitch-finding algorithm, used to construct $p(\Delta\Omega)$, is described in Sec.~\ref{subsec:glitchfind}.  \emph{Top left}:  Angular velocity as a function of time, $\Omega_{\rm{c}}(t)$, for the time interval $2.519<t<2.557$ (\emph{solid} curve).  The \emph{dotted} curve is a linear fit to $\Omega_{\rm{c}}(t)$ with slope $-0.390$.  \emph{Top right}:  PDF of jumps $\Delta\Omega_{\rm{c}}$ for three glitch-finding time windows, $\Delta t = 10^{-2.0}$, $10^{-1.5}$ and $10^{-1.0}$.  \emph{Centre left}:  Time series of waiting times since previous glitch, $\tau$, versus glitch epoch $t$. \emph{Centre right}:  PDF of $\tau$ on log-linear axes (\emph{solid} curve).  The \emph{dotted} curve is an exponential least-squares fit with exponent $\theta = -0.42$.  \emph{Bottom left}:  Shear parameter as a function of time, $\gamma(t)$.  \emph{Bottom right}:  PDF of $\gamma$.  The \emph{dashed vertical} line indicates the position of $\gamma_{\rm{eq}}=0.923$, which is offset from the mean $\langle\gamma \rangle=0.922$.}
\label{fig:basic}
\end{figure}

A glitch brings the coupled superfluid-crust system closer to corotation.  Immediately after the event, the superfluid decelerates and the crust accelerates, reducing the vortex unpinning rate by increasing $\gamma$ [see Eq.~(\ref{eq:poftau})].  After a longer delay, the electromagnetic torque decelerates the crust (without affecting the superfluid), gradually reducing $\gamma$ and hence increasing $\theta$.  Spin-up of the crust in response to downward steps in superfluid rotation ensures that the glitch mechanism is self-renewing.  That is, each time a glitch occurs, the lag between the crust and superfluid is partially reset (almost never to zero)\footnote{Since no correlation is observed between glitch size and waiting time, it can be deduced that the lag between the superfluid and crust rotation is not reduced to zero by a glitch.}, reinitiating the accumulation of lag that leads to the glitch.   Without feedback, the lag would continue to grow, eventually rendering pinning impossible.  On very long time-scales ($\gtrsim 10^6\,\rm{yr}$), therefore, the system is not stationary.  However, the slow electromagnetic spin-down rate, $d\Omega_{\rm{c}}/dt$ ($\sim10^{-13}\,\rm{Hz\,s}^{-1}$, for a typical pulsar) suggests that, over the forty years of historical pulsar observations we can treat $\Omega_{\rm{c}}$ as constant.  

The crust responds instantaneously to changes in the superfluid angular velocity according to
\begin{equation}
\label{eq:feedback}
 I_{\rm{c}}\frac{d\Omega_{\rm{c}}}{dt}=-I_{\rm{s}}\frac{d\Omega_{\rm{s}}}{dt}+N_{\rm{c}}~,
\end{equation}
where $I_{\rm{c}}$ and $I_{\rm{s}}$ are the moment of inertia of the pulsar crust [and the charged proton and electron fluids that are tightly coupled to it \citep{Alpar:1984p6781}] and the superfluid (notionally, if it were a rigid body of the same density) respectively, and $N_{\rm{c}}$ is the electromagnetic torque on the pulsar crust \footnote{We remind the reader that the observed deceleration of a pulsar is a compromise between $N_{\rm{c}}$ and the time-averaged spin-up torque exerted by the decelerating superfluid, which is composed of both smooth and discrete jumps.  Therefore, we cannot merely write $N_{\rm{c}}=I_{\rm{c}}\dot{\Omega}_{\rm{c}}$.}.  In practice, the time scale on which changes in the superfluid angular momentum are communicated to the crust is governed by the crossing time of a Kelvin wave along the length of a vortex [$\sim 10^{-1}\,\rm{s}$ \citep{Epstein:1992p2172}].  We can rewrite Eq.~(\ref{eq:feedback}) in terms of changes to the global shear parameter $\gamma$,
\begin{eqnarray}
  \frac{d\gamma}{dt}&=&\frac{1}{\Delta\Omega_{\rm{cr}}}\left[-\left(1+\frac{I_{\rm{s}}}{I_{\rm{c}}}\right)\frac{d\Omega_{\rm{s}}}{dt}+\frac{N_{\rm{c}}}{I_{\rm{c}}}\right]~.
\end{eqnarray}
It should be noted, that interactions between the viscous component of the internal fluid (composed predominantly of protons and electrons) and the neutron superfluid also result in a coupling to the crust, hence influencing the speed and strength of coupling.  The likely time scale of such an interaction is the Kelvin wave crossing time of the star, which is also equivalent to the mutual friction coupling time \citep{Mendell:1991}, which has been studied recently in a hydrodynamical context in glitch recovery models \citep{VanEysden:2010}.

Strictly speaking, a constant-rate ($d\gamma/dt=0$, constant $\beta$, $E_{\rm{p}}$ and $\Gamma_0$) Poisson process does not capture the self-organising behaviour that is essential to regulating the crust-superfluid lag, since changes in the lag are equivalent to changes in the unpinning rate.  Fluctuations in $\gamma$, caused by glitches, render vortex unpinning an inherently non-constant-rate process.  The absence of a reservoir effect in glitch data (glitch sizes are not correlated with waiting times) indicates that glitches never completely nullify the lag, so the stress released in each glitch is a combination of recent and historical build up \citep{Melatos:2008p204}.  In Sec.~\ref{sec:master} we discuss an analytic formalism for describing a variable rate, or \emph{state-dependent}, Poisson process.  For now, however, we investigate the behaviour of a system with a constant rate, employing Eq.~(\ref{eq:feedback}) to ensure that unpinning balances the spin down of the crust on average over the long term.

The change in angular momentum of the superfluid when one vortex unpins and moves outward a fraction $\alpha$ of the stellar radius is given by (see Appendix~\ref{app:dL})
\begin{equation}
\Delta L = 2\pi\rho\kappa R^3 \alpha~.
\end{equation}
Gross-Pitaevskii simulations suggest that $\alpha R$ approximately equals the mean inter-vortex spacing \citep{Warszawski:2010individual}, with
\begin{eqnarray}
 \alpha&=&  \frac{1}{R}\left(\frac{\kappa}{2\Omega}\right)^{1/2}~,
\end{eqnarray}
whether the number of vortices exceeds or is exceeded by the number of pinning sites \citep{Warszawski:2010pulsar}.
In order to find the asymptotic spin-down rate, $d\Omega_{\rm{s}}/dt$, we now consider a constant-rate process, with $\theta[E_{\rm{p}},\gamma(t)]$ defined in Eq.~(\ref{eq:rate}).  That is, for $d\gamma/dt=0$ we get
\begin{equation}
\label{eq:fb}
 \frac{d\Omega_{\rm{s}}}{dt}=\frac{N_{\rm{c}}/I_{\rm{c}}}{1+I_{\rm{s}}/I_{\rm{c}}}~.
\end{equation}
Although in practice $d\gamma/dt$ is non-zero, self-regulation, via Eq.~(\ref{eq:feedback}) ensures that the unpinning rate results in the commensurate spin down of the superfluid with the crust.  On average, the superfluid spins down at a rate proportional to the mean vortex unpinning rate, viz.
\begin{eqnarray}
\label{eq:domdtavg}
 \frac{d\Omega_{\rm{s}}}{dt}
 &=&\frac{\Delta L}{I_{\rm{s}}}N_{\rm{v}}\Gamma_0 \exp\left(-\beta E_{\rm{p}}\gamma_{\rm{eq}}\right)~.
\end{eqnarray}
Combining Eq.~(\ref{eq:fb}) and (\ref{eq:domdtavg}), and solving for $\gamma$, we have an estimate for the steady-state shear parameter, $\gamma_{\rm{eq}}$,
\begin{equation}
\label{eq:eqgamsimple}
 \gamma_{\rm{eq}}=-\frac{1}{\beta E_{\rm{p}}}\ln\left(\frac{N_{\rm{c}}}{\Gamma_0 N_{\rm{v}}\Delta L}\frac{I_{\rm{s}}/I_{\rm{c}}}{1+I_{\rm{s}}/I_{\rm{c}}}\right)~,
\end{equation}
which ensures that superfluid spin down by unpinning matches the rate dictated by the electromagnetic torque.

The \emph{left} panel of Fig.~\ref{fig:schematic} illustrates how $\Omega_{\rm{c}}$ and $\Omega_{\rm{s}}$ change when allowed to evolve according to Eq.~(\ref{eq:feedback}) for a set of artificially generated, power-law-distributed glitch sizes, imposed at exponentially distributed intervals on a background of constant deceleration.  Between unpinning events, $\Omega_{\rm{s}}$ (\emph{grey} curve) does not change, whilst $\Omega_{\rm{c}}$ (\emph{black} curve) decreases linearly with time at a rate $N_{\rm{c}}/I_{\rm{c}}$.  When vortices unpin, the $\Omega_{\rm{s}}$ curve steps down instantaneously, matched by an instantaneous acceleration of the crust.  In the \emph{right} panel we graph $\gamma(t)$ over the same time interval; between unpinning events, $\gamma(t)$ decreases linearly at a rate $N_{\rm{c}}/(I_{\rm{c}}\Delta\Omega_{\rm{cr}})$.  It increases instantaneously  by $(\Delta\Omega-\Delta\Omega_{\rm{s}})/\Delta\Omega_{\rm{cr}}$ when a vortex unpins, with $\Delta\Omega = \Delta L/I_{\rm{c}}$.

\begin{figure}
\includegraphics[scale=0.36,angle=90]{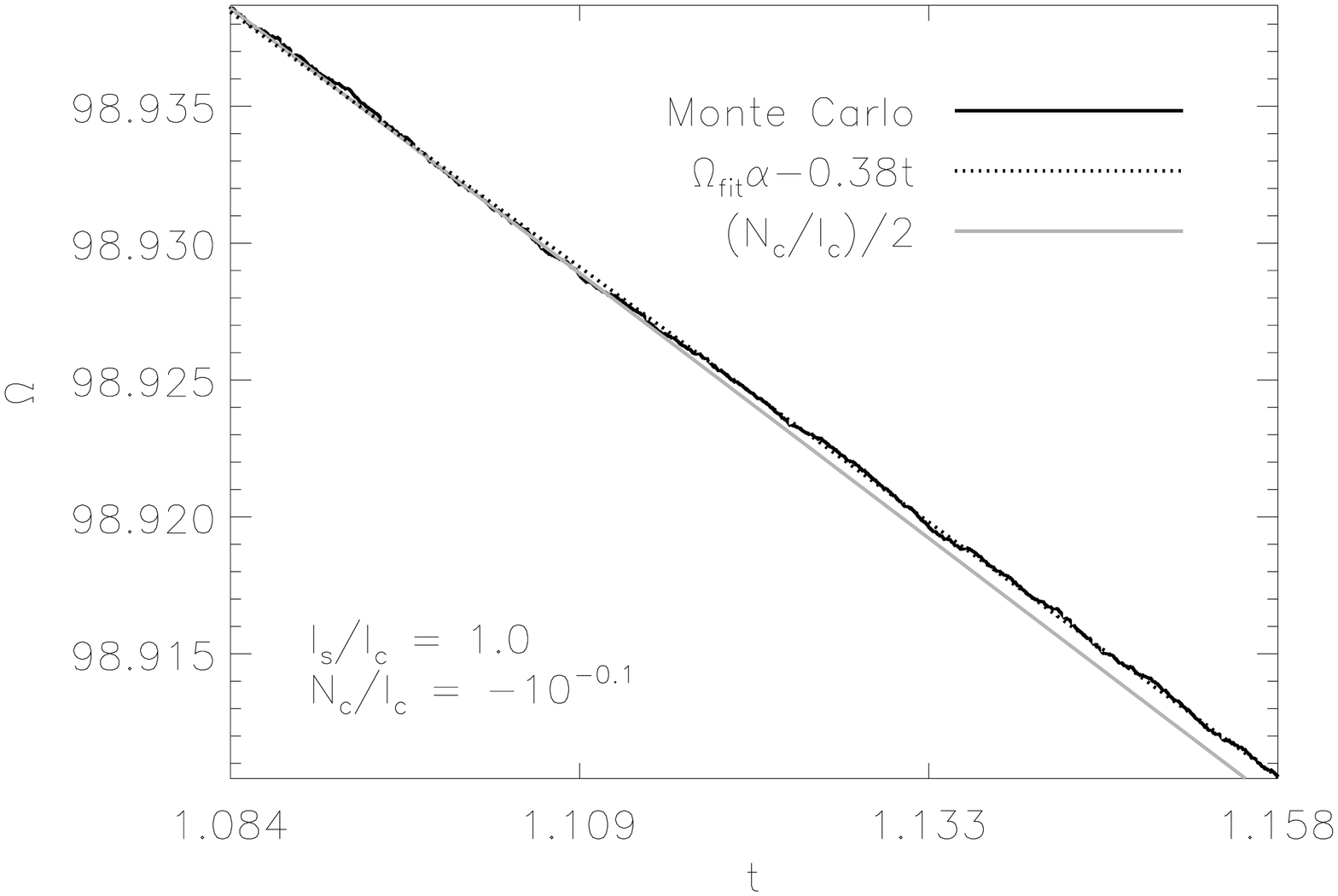}
\includegraphics[scale=0.36,angle=90]{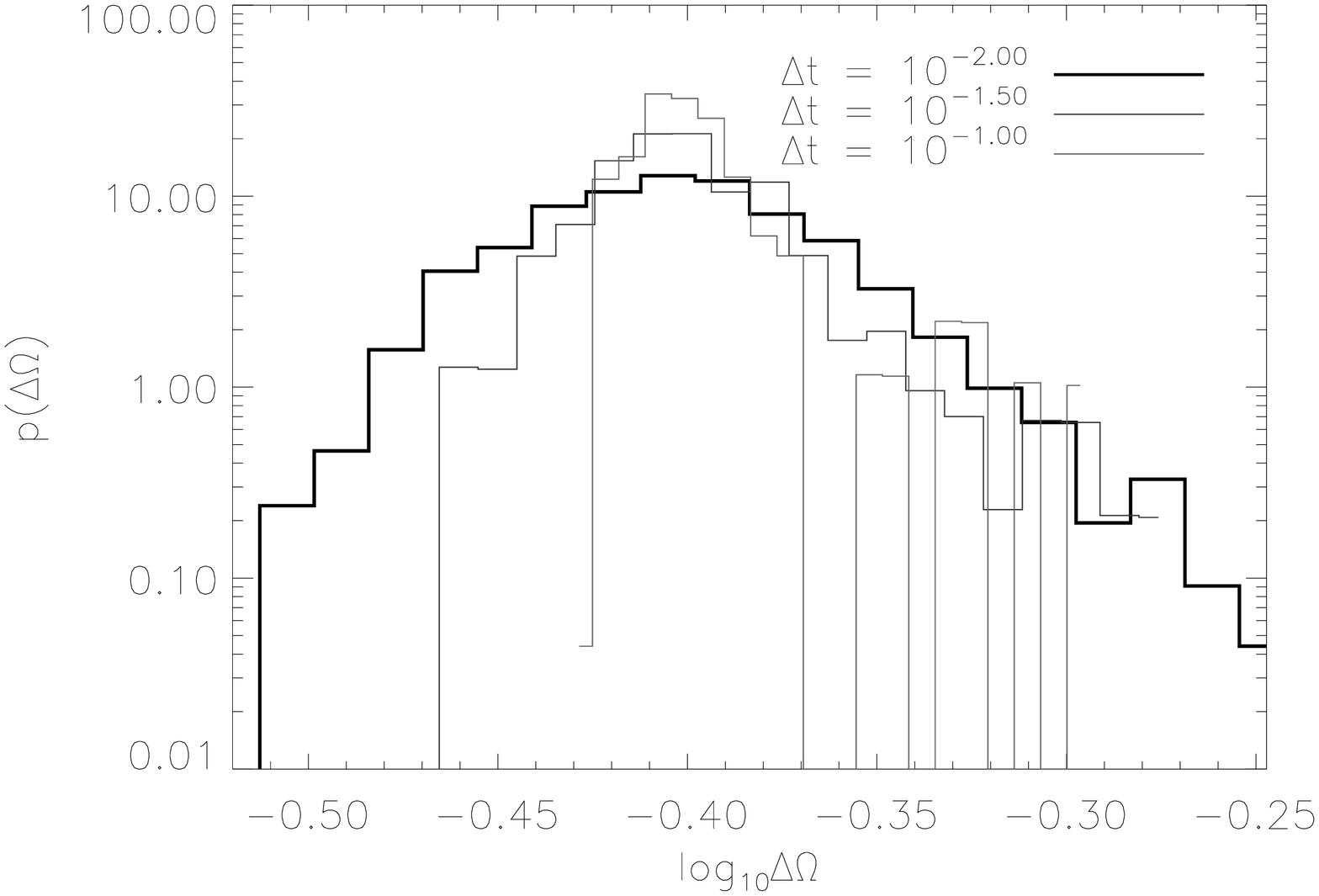}
\includegraphics[scale=0.36,angle=90]{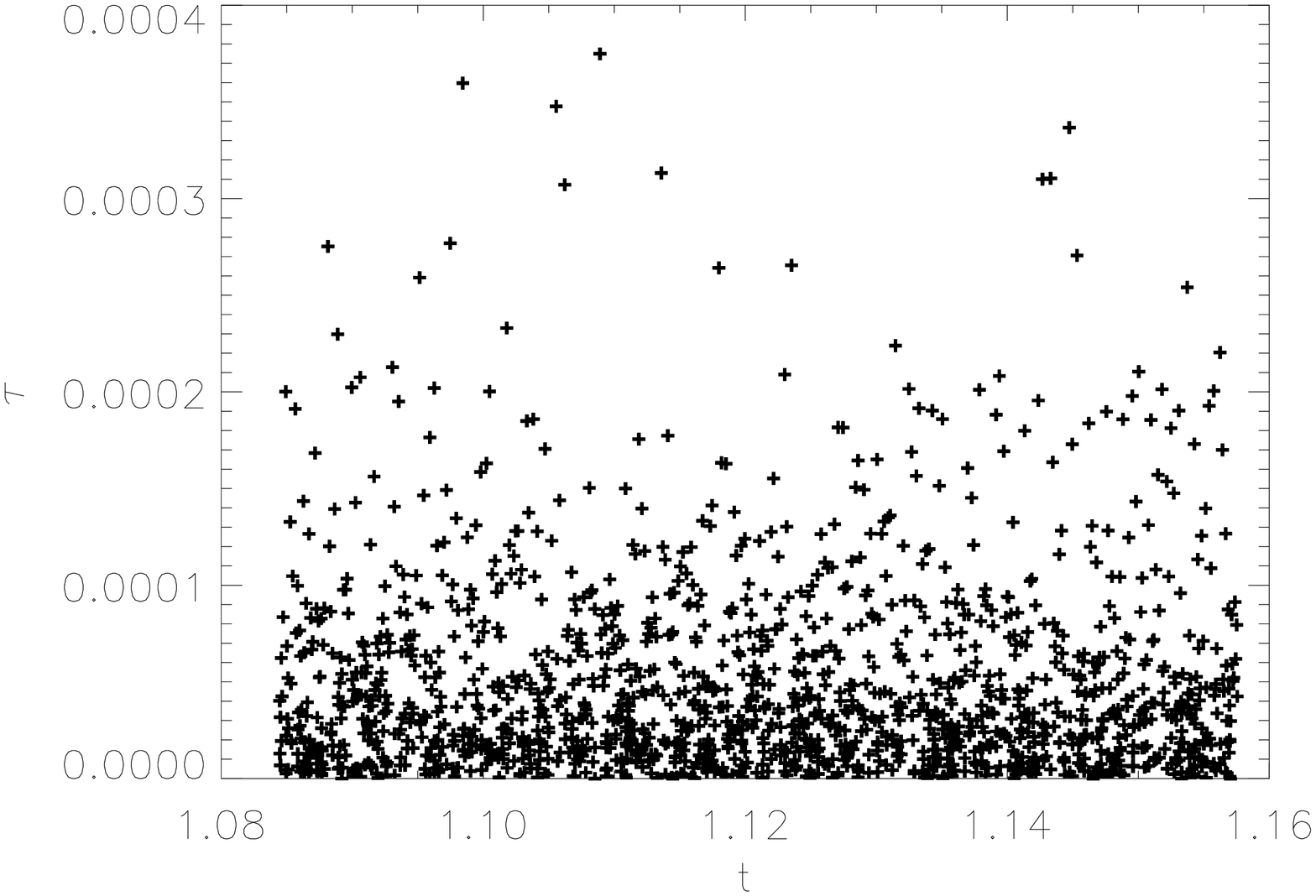}
\includegraphics[scale=0.36,angle=90]{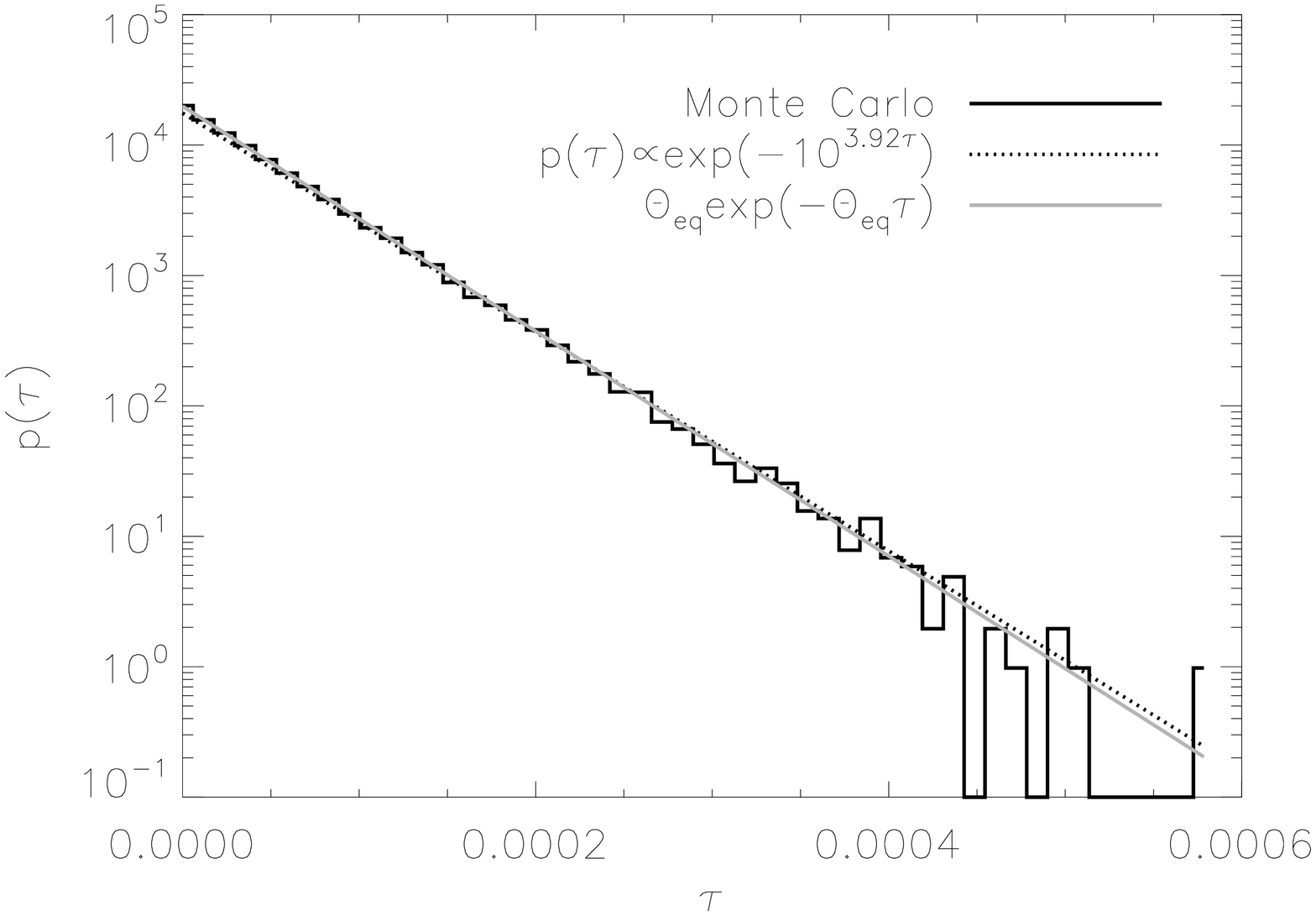}
\includegraphics[scale=0.36,angle=90]{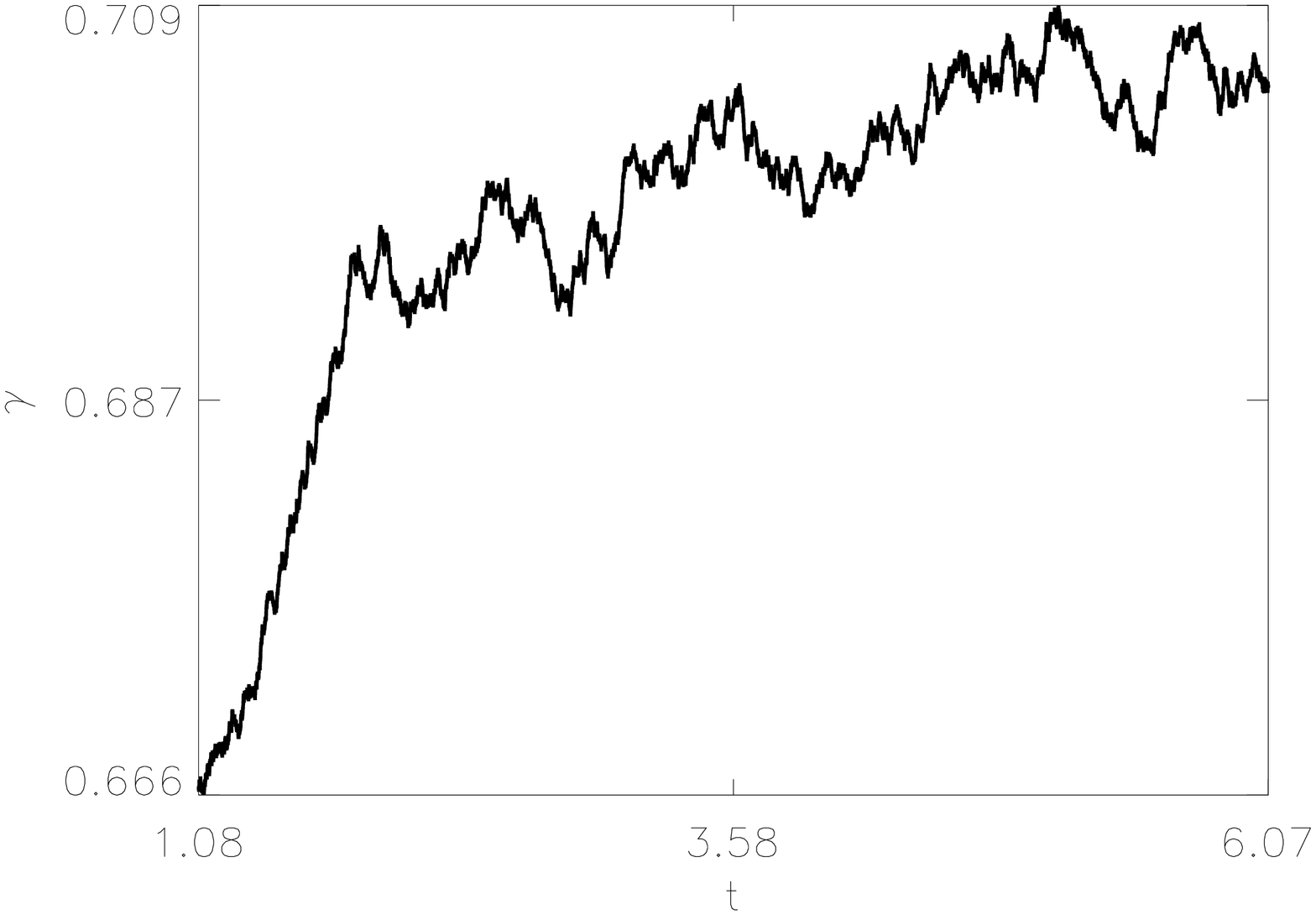}
\includegraphics[scale=0.36,angle=90]{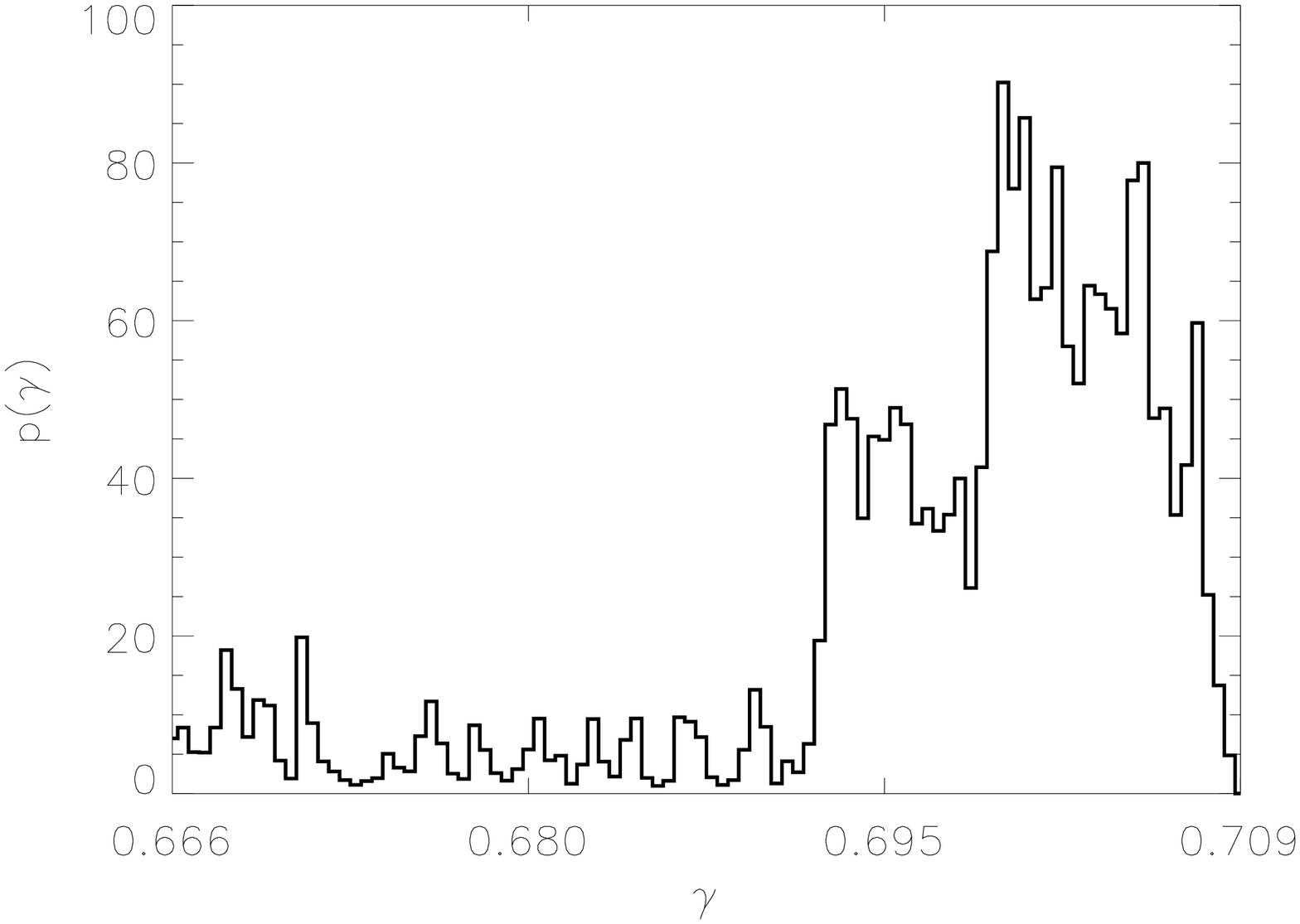}
\caption{Monte-Carlo simulations for the vortex system in Fig.~\ref{fig:basic}, but with a range of available pinning energies  $E_{\rm{p}}\in[E_0-\Delta E_{\rm{p}},E_0+\Delta E_{\rm{p}}]$.  The automaton rules are described in Sec.~\ref{sec:MCsimple}.  Results are reported as time series and PDFs.  \emph{Top left}:  Angular velocity as a function of time, $\Omega_{\rm{c}}(t)$, for the time interval $1.084<t<1.158$ (\emph{solid} curve).  The \emph{dotted} curve is a linear fit to $\Omega_{\rm{c}}(t)$ with slope $-0.390$.  \emph{Top right}:  PDF of jumps, $\Delta\Omega_{\rm{c}}$.  \emph{Centre left}:  Time series of time intervals between unpinning events, $\tau$. \emph{Centre right}:  PDF of $\tau$ on log-linear axes (\emph{solid} curve).  The \emph{dotted} curve is an exponential least-squares fit with rate parameter $\theta = -0.42$.  \emph{Bottom left}:  Shear parameter as a function of time $\gamma(t)$.  \emph{Bottom right}:  PDF of $\gamma$. $\gamma_{\rm{eq}}=0.751$ falls outside the plotted range; it is larger than the average $\langle\gamma\rangle=0.697$.}
\label{fig:basicdE}
\end{figure}

\section{Monte-Carlo simulations}\label{sec:MCsimple}
In the remainder of the paper, we implement the physics in Sec.~\ref{sec:unpinningknock} and \ref{sec:feedback} in an automaton to study the open questions described in \S~{II}.  We vary the rules governing the vortex unpinning rate and the distribution of pinning energies and analyse the output statistically, aided by a glitch-finding algorithm, which compiles glitch waiting-time and size distributions (see Sec.~\ref{subsec:glitchfind}) .
 
The automaton output and the analytic estimates given in Sec.~\ref{sec:feedback} are compared against $\gamma_{\rm{eq}}$, given in Eq.~(\ref{eq:eqgamsimple}), and the mean of the simulated $\gamma$ distribution.  Equation~(\ref{eq:eqgamsimple}) only applies for uniform pinning energies, $E_{\rm{p}}=E_0$.  Later, in Sec.~\ref{sec:master}, we write down a master equation [Eq.~(\ref{eq:CK})] that corresponds to the uniform-pinning scenario.  We extend the simulations and theory to encompass a range of $E_{\rm{p}}$ in the next section.

\subsection{Asynchronous automaton}\label{subsec:asynch}

In an asynchronous cellular automaton, the time step is a stochastic variable that depends on the state of the system \footnote{The `best choice' for the order in which cells are updated depends on the specific application; it has been shown that the choice of update scheme can greatly influence the simulation output \citep{Cornforth:2005}.}.

Each step in the automaton is governed by the following rules.
\begin{enumerate}
 \item For each vortex $i$, a time $\tau_{i}$ until the next unpinning is drawn from $p(\tau;E_{\rm{p}},t)$ [defined in Eq.~(\ref{eq:poftau})], with the initial condition  $\gamma(t=0)=\gamma_0$.
 \item The system is advanced in time by $\tau_{\rm{min}}=\rm{min}\{\tau_{i}\}$, and the global time counter becomes $t_{n}=t_{n-1}+\tau_{\rm{min}}$ \citep{Durrett:1995}.
 \item $\Omega_s$ is decremented by $\Delta\Omega_s=\Delta L/I_{\rm{s}}$, and $\Omega_c$ changes by $\Delta\Omega_c=-\Delta L/I_{\rm{c}}+\tau_{\rm{min}}N_{\rm{c}}/I_{\rm{c}}$ [see Eq.~(\ref{eq:feedback})].  The shear parameter becomes $\gamma(t+\tau_{\rm{min}})=\gamma(t)+(\Delta\Omega_c-\Delta\Omega_s)/\Delta\Omega_{\rm{cr}}$.
 \item The unpinned vortex is assigned a new waiting time (and a new pinning strength if $\Delta E_{\rm{p}}\neq 0$), drawn from $p(\tau;E_{\rm{p}},t)$.
 \item For all vortices that did not unpin in the previous interval, the time until the next unpinning event becomes \begin{equation}\label{eq:stretch}
     (\tau_{i}-\tau_{\rm{min}})\exp\left[-\beta E_0 \gamma(t+\tau_{\rm{min}})\right]/\exp\left[-\beta E_0 \gamma(t)\right]~. \end{equation}
 \item Repeat steps 2--5.
\end{enumerate}
Rules 4 and 5 define what is known in the literature as a time-driven asynchronous automaton; instead of updating every cell at each time step, cells are updated in the order that Poisson events affecting them actually occur.  The stationary states of an automaton do not depend on whether the update rules are synchronous or asynchronous, but this is not true of their basins of attraction \citep{Huberman:1993,Schonfisch:1999}.  Asynchronous update rules are both efficient and flexible.  When the event rate is slow, an asynchronous algorithm avoids unnecessarily small time steps, while the lack of a minimum time step means there is no upper bound on event rates.  

Rule 5 is necessary because each event changes the state of the system and hence the unpinning rate for all vortices.  Equation~(\ref{eq:stretch}) encodes the response of the unpinning rate to the new state of the system, by stretching the remaining fraction of the waiting time according to the ratio of new and old rates.  That is, the vortex with the shortest waiting time unpins after $\tau_{\rm{min}}$ time units.  Another vortex $j$ would have unpinned in $\tau_{j}-\tau_{\rm{min}}$ later without any adjustment.  However for the remainder of the waiting time, the unpinning rate changes and, to compensate, we stretch $\tau_{j}-\tau_{\rm{min}}$ by the ratio of the new and old unpinning rates, $\Gamma_0\exp[-\beta E_0\gamma(t+\tau_{\rm{min}})]$ and $\Gamma_0\exp[-\beta E_0\gamma(t)]$ respectively.

An alternative approach would be to reset the waiting times of all vortices at the beginning of each new time interval.  Such an approach agrees with the spirit of a Poisson process, in which time intervals of any length are independent, but it is not favoured by other authors.  We follow the majority custom here.  

In all forthcoming plots, quantities are given in arbitrary units.

\subsection{Glitch classification}\label{subsec:glitchfind}

A robust, automated glitch-finding algorithm is essential for compiling and studying glitch statistics.  The simplest approach is to calculate the accumulated spin up, $\Delta \Omega_{\rm{c}}$, during some time window of width $\Delta t$, and then construct the probability density function (PDF) of $\Delta\Omega_{\rm{c}}$, $p(\Delta \Omega_{\rm{c}})$, by sliding this window along the time series $\Omega_{\rm{c}}(t)$.   

The shape of $p( \Delta \Omega_{\rm{c}})$ depends strongly on $\Delta t$.  Figure~\ref{fig:mock} graphs $\Omega_{\rm{c}}(t)$ for a mock system, in which glitches with power-law-distributed sizes (exponent $a=-3/2$) and exponentially distributed waiting times are superposed on linear deceleration, viz.,
\begin{equation}
 \Omega_{\rm{c}}(t)=\Omega_{\rm{c}}(0)+\dot{\Omega}_{\rm{c}}t+\sum_{t \in {t_{\rm{g}}}}\Delta\Omega_{\rm{g}}(t)~,
\end{equation}
where $\Delta\Omega_{\rm{g}}(t)$ represents the intermittent glitches and is non-zero only when a glitch occurs at $t=t_{\rm{g}}$ (in arbitrary units).  The \emph{left} panel graphs $\Omega_{\rm{c}}(t)$.  The \emph{right} panel graphs $p( \Delta \Omega_{\rm{c}})$ for $\Delta t=10^{-1.75}$, $10^{-1.25}$, $10^{-0.75}$, $10^{-0.25}$ (\emph{dark} to \emph{light} curves respectively).  For the two smallest values of $\Delta t$, $p(\Delta\Omega_{\rm{c}})$ is well represented by a power law, but the best-fit power-law index ($a = -1.31$ for $\Delta t=10^{-1.75}$) does not match exactly the distribution from which the glitches were drawn ($a=-3/2$).  For larger values of $\Delta t$, $p( \Delta \Omega_{\rm{c}})$ peaks about a mean value. This behaviour is a direct result of the central limit theorem.  When the time window is much larger than the mean inter-glitch waiting time, each window brackets several glitches, and hence $\Delta\Omega_{\rm{c}}$, which is the sum of several draws from the same underlying size PDF, populates a Gaussian distribution.  Therefore, we conclude that a narrow window with $\Delta t\lesssim \theta^{-1}$ best characterises the underlying glitch behaviour.

\section{Pinning at a unique energy}\label{subsec:homoopin}

The simplest possible form of the vortex (un)pinning theory of glitches treats the pinning strength as uniform throughout the pulsar.  Naively, this scenario should produce quasi-periodic glitches with a narrow range of sizes, since each vortex sustains roughly the same level of stress before unpinning.  In this section we test this idea using the automaton described in Sec.~\ref{sec:MCsimple}.

Results from Monte-Carlo simulations for single-energy pinning ($E_0=1$, $\Delta E_{\rm{p}}=0$) are shown in Fig.~\ref{fig:basic}.  The electromagnetic torque is $N_{\rm{c}}=10^{-0.1}I_{\rm{c}}$, each unpinned vortex travels a distance $\alpha R=10^{-4}R$, the crust and superfluid moments of inertia are taken to be equal, ($I_{\rm{s}}/I_{\rm{c}}=1$), and there are $N_{\rm{v}}=10^4$ vortices.  The rules are enacted $N_{\rm{t}}=10^7$ times.  The graphs show the angular velocity of the crust as a function of time $\Omega_{\rm{c}}(t)$, and its PDF, $p(\Delta\Omega_{c})$, waiting times between unpinning events $\tau$ as a function of time, and their PDF $p(\tau)$, $\gamma$ as a function time, and its PDF $p(\gamma)$ (from \emph{top} to \emph{bottom}, \emph{left} to \emph{right}).  The PDFs are constructed by sampling the quantity at each time step and constructing a histogram of the sampled values.  Calculation of $p(\Delta\Omega_{\rm{c}})$ is discussed in Sec.~\ref{subsec:glitchfind}.

The time series of $\Omega_{\rm{c}}$ exhibits restless behaviour, including spin-up events.  The \emph{dotted} curve is a straight line with slope $-N_{\rm{c}}/(2I_{\rm{c}})$, which is the expected spin-down rate for $\gamma=\gamma_{\rm{eq}}$ and $I_{\rm{s}}/I_{\rm{c}}=1$.  Although the simulated $\Omega_{\rm{c}}$ wanders away from the \emph{grey} curve, $-N_{\rm{c}}/(2I_{\rm{c}})$ remains a good approximation for the global spin-down rate over long time periods.  The size distribution, $p(\Delta\Omega_{c})$, is strongly peaked, with mean and standard deviation 0.40 and 0.02 respectively.  It does not span many decades, unlike real observed glitches.  Further evidence that fluctuations in $\Omega_{\rm{c}}$ due to glitches do not push it far from the time-averaged trend is found in the $p(\gamma)$ curve, for which $\gamma_{\rm{eq}}=0.923$ [\emph{dashed} vertical line, taken from Eq.~(\ref{eq:eqgam})] lies within the narrow range of values (mean and standard deviation 0.92 and 0.0023 respectively) visited by the automaton.  However, we note that $\gamma_{\rm{eq}}$ does not coincide exactly with the mode of $p(\gamma)$.

The narrow distribution of $\gamma$ values leads naturally to an almost exponential distribution of waiting times because the activity approaches a constant-rate Poisson process, with mean rate given by $\theta \approx N_{\rm{v}}\Gamma_0\exp(-\beta\gamma_{\rm{eq}}E_0)$ (\emph{grey} curve in \emph{middle left} panels of Fig.~\ref{fig:basic} and \ref{fig:basicdE}).  The \emph{dotted} curve is an exponential least squares fit, whose slope differs from $-\beta\gamma_{\rm{eq}}E_0$ by $0.053$.

We conclude that a state-independent Poisson process captures well the behaviour of the single-pinning-energy system.  However, such a system does not give rise to a broad distribution of event sizes, like those observed astronomically.  In the following sections, we repeat this analysis for systems with (i) a wide range of pinning energies, and (ii) collective triggers for unpinning avalanches (Sec.~\ref{sec:heteropinning} and \ref{sec:branching} respectively). 

\begin{figure}
\begin{center}
\includegraphics[scale=0.45]{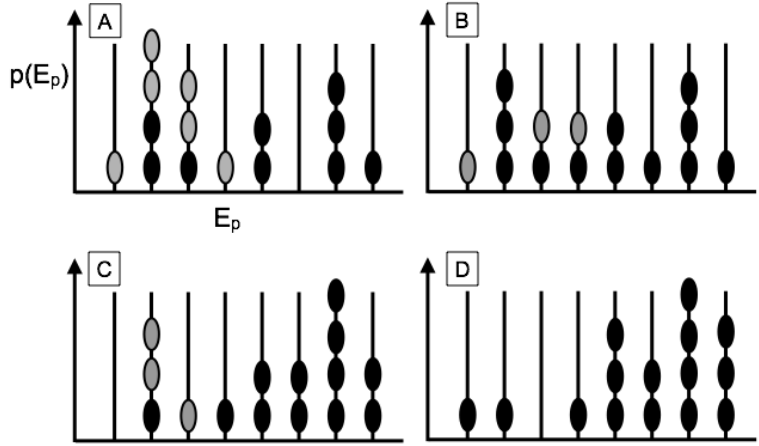}
\end{center}
\caption{Schematic of how vortices accumulate over time at strong pinning sites, biasing the distribution of occupied pinning sites, $g(E_{\rm{p}})$, towards large $E_{\rm{p}}$.  Since weakly pinned vortices repin more frequently than strongly pinned vortices, and a vortex repins with a strength chosen at random from a uniform distribution, $\phi(E_{\rm{p}})$, over time, weak pinning sites are depopulated in favour of strong pinning sites.  The panels depict this process taking place over four episodes, starting with uniformly-distributed, random pinning strengths (\emph{top left}), and ending with a preference for occupying strong pinning sites (\emph{bottom right}).  In each panel, the vortices which are about to unpin (and repin uniformly in the next panel) are coloured grey.}
\label{fig:Ehist_diag}
\end{figure}
\begin{figure}
\begin{center}
\includegraphics[scale=0.4,angle=90]{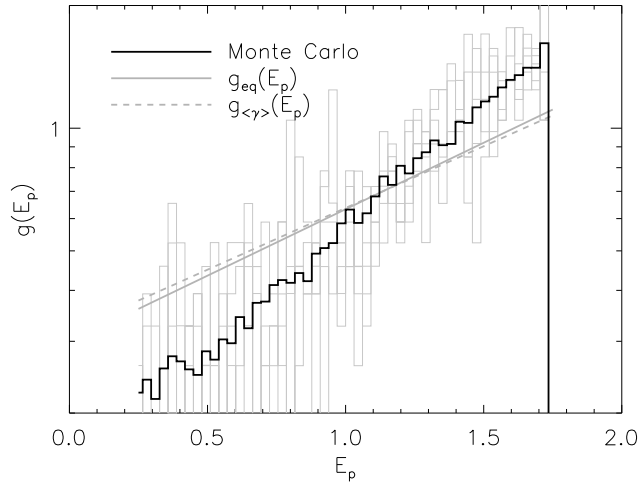}
\end{center}
\caption{Pinning energy distribution of occupied sites, $g(E_{\rm{p}})$, for the Monte-Carlo simulation reported in Fig.~\ref{fig:basicdE} (\emph{solid} curve), overplotted with $g_{\rm{eq}}(E_{\rm{p}})$ from Eq.~(\ref{eq:knock8}) (\emph{dashed} curve).  The \emph{solid grey} curve is $g(E_{\rm{p}})$, calculated using $\gamma = \langle\gamma\rangle$, taken from $p(\gamma)$ (\emph{bottom right} panel of Fig.~\ref{fig:basicdE}).  The \emph{light grey} histograms display the instantaneous $g(E_{\rm{p}},t)$, at five equally spaced times.}
\label{fig:Ehist}
\end{figure}
\section{Pinning at multiple energies}\label{sec:heteropinning}

The strength of pinning within a pulsar is likely to be heterogeneous.  Small-scale (microscopic) heterogeneities arise from defects in the crystalline lattice of the crust, in the form of either vacancies, dislocations, or impurities \citep{Pizzochero:1997p45,DeBlasio:1998p127,Donati:2003p97,Pizzochero:2007p144}.  Large-scale (macroscopic) fluctuations in the pinning strength may result from grain boundaries in the crust, creating reservoirs of strongly pinned vortices \citep{Alpar:1984p6781,CHENG:1988p180}.  We approximate the effect of small-scale variations by choosing the pinning energy at each pinning site from a distribution function $\phi(E_{\rm{p}})$, defined as
\begin{equation}
\label{eq:phiofE}
 \phi(E_{\rm{p}})= (2E_0)^{-1} H(E_0-\Delta E_{\rm{p}} - E_{\rm{p}}) H(E_0+\Delta E_{\rm{p}} - E_{\rm{p}})~,
\end{equation}
where $E_0$ is the mean pinning energy, $\Delta E_{\rm{p}}$ is the half-width of the distribution, and $H(\cdot)$ is the Heaviside step function.  Whenever a vortex repins, its new pinning energy is chosen  at random from $\phi(E_{\rm{p}})$.

Heterogeneous pinning introduces a new statistical quantity, namely the distribution of \emph{occupied} pinning energies, $g(E_{\rm{p}},t)$.  This is not the same as $\phi(E_{\rm{p}})$ \citep{Newman:1996p1484,Melatos:2008p204}; shallow pinning wells are less likely to be occupied than deep ones, in a statistical sense.  In Sec.~\ref{subsec:excavate} we estimate $g(E_{\rm{p}},t)$ for a stationary system analytically.

\subsection{Automaton output}\label{subsec:heteroauto}
Figure~\ref{fig:basicdE} graphs results from Monte Carlo simulations for heterogeneous pinning ($\Delta E_{\rm{p}}/E_0=0.5$, $E_0=1$, $I_{\rm{s}}/I_{\rm{c}}=1$, $N_{\rm{c}}/I_{\rm{c}}=10^{-0.1}$, $\alpha=10^{-4}$, $N_{\rm{v}}=10^4$, $N_{\rm{t}}=10^7$).  The graphs show the angular velocity of the crust as a function of time $\Omega_{\rm{c}}(t)$, the PDF of spin-up events, $p(\Delta\Omega_{c})$, waiting times between unpinning events $\tau$ as a function of glitch epoch, and the PDF $p(\tau)$, and $\gamma$ as a function of time and the PDF (from \emph{top} to \emph{bottom}, \emph{left} to \emph{right}).  

Qualitatively, the spin-down curve (\emph{top left} panel) does not differ greatly from the unique pinning energy case (\emph{top left} panel of Fig.~\ref{fig:basic}).  The PDF of $\Delta\Omega_{\rm{c}}$ confirms that heterogeneous pinning strengths are insufficient to significantly broaden (\emph{i.e}. across several decades) the $\Delta\Omega_{\rm{c}}$ distribution; the range of $\Delta\Omega_{\rm{c}}$ is less than one decade ($10^{-0.6}<\Delta\Omega_{\rm{c}}<10^{0.2}$), slightly but not much wider than $p(\Delta\Omega_{\rm{c}})$ in Fig.~\ref{fig:basic}.

The validity of a state-independent ($\gamma=\gamma_{\rm{eq}}$) description of the system is confirmed by the relatively narrow distribution of $\gamma$ ($0.79\lesssim\gamma\lesssim 0.8$).  The waiting times  (\emph{middle right} panel) are well represented by an exponential distribution with mean rate given by the equilibrium unpinning rate, $\theta_{\rm{eq}}=\theta(E_{\rm{p}},\gamma_{\rm{eq}})$ (\emph{solid grey} curve), which in turn closely matches the exponential fit plotted as a \emph{dotted} curve.

\subsection{Excavation}\label{subsec:excavate}

Vortices unpin readily from weak pinning sites, and with difficulty from strong pinning sites, as illustrated schematically in Fig.~\ref{fig:Ehist_diag} (the sequence is from A---D).  This tendency `excavates' the PDF of occupied energies so that the majority of vortices sit on strong pinning sites.  In the initial state, immediately following a system-spanning glitch, vortices are pinned at sites with uniformly distributed energies (as in frame A). Vortices that subsequently unpin are coloured \emph{grey}.  Weakly pinned vortices are more likely to unpin and repin at randomly distributed energies (as in frame B).  A preference for strongly pinned vortices emerges (frame D).  This behaviour, to which broadly distributed glitch sizes were attributed, was also observed in the coherent noise model described in \cite{Melatos:2009p4511}.

Let $g(E_{\rm{p}},t)dE_{\rm{p}}$ be the fraction of vortices at time $t$ pinned at sites with pinning energies in the range $(E_{\rm{p}},E_{\rm{p}}+dE_{\rm{p}})$.  In a time $d t$, in the energy range $(E_{\rm{p}},E_{\rm{p}}+dE_{\rm{p}})$, a number of vortices equal to $\theta(E_{\rm{p}},\gamma) g(E_{\rm{p}},t) dt$ unpin.   Some repin in the same range $(E_{\rm{p}},E_{\rm{p}}+dE_{\rm{p}})$, in numbers proportional to the fraction $\phi(E_{\rm{p}})dE$ of available sites in that range, and some repin at other sites with energies outside this range. Inspired by models of atomic hopping in glasses \citep{Bouchaud:1995,Rinn:2001}, this stochastic process can be described statistically via a master equation \citep{Monthus:1996p10195,Head:2000}
\begin{equation}
 \frac{\partial g(E_{\rm{p}},t)}{\partial t} = -\theta[E_{\rm{p}},\gamma(t)] g(E_{\rm{p}},t) + \omega(t) \phi(E_{\rm{p}})~,
\label{eq:masterg}
\end{equation}
where $\omega(t)$ is the unpinning rate at time $t$ averaged over all occupied sites,
\begin{equation}
\label{eq:rateeq}
 \omega(t) = \int_0^\infty dE_{\rm{p}}' \,\theta[E_{\rm{p}}',\gamma(t)]g(E_{\rm{p}}',t)~.
\end{equation}
Equation (\ref{eq:masterg}) is solved supplemented by the initial condition $g(E_{\rm{p}},0)=\phi(E_{\rm{p}})$ \citep{Bak:1993} 
by taking the Laplace transform (see Appendix~\ref{sec:knockappa} ).  We know from observations --- e.g.\ the universal exponential waiting-time distributions observed in pulsars on the time-scale of years \citep{Melatos:2008p204} --- that the glitch statistics and therefore presumably the underlying $g(E_{\rm{p}},t)$ reach stationarity fairly quickly.  From Eq.~(\ref{eq:masterg}), in the steady state [$\partial g(E_{\rm{p}},t)/\partial t=0$], $\omega(t)$ must tend to a constant $\omega_{\rm eq}(\beta,\gamma_{\rm{eq}})$, with 
\begin{eqnarray} 
\label{eq:knock5}
\omega_{\rm eq}
 & = &
 \left[\frac{1}{N_{\rm{v}}\Gamma_0}
 \int_0^\infty dE' \, e^{\beta\gamma_{\rm{eq}} E'} \phi(E')
 \right]^{-1}~,\\
 & = &
 \frac{2\Delta E_{\rm{p}}\beta \gamma_{\rm{eq}}e^{-\beta\gamma_{\rm{eq}} E_0}}{e^{\beta\gamma_{\rm{eq}}\Delta E_{\rm{p}}} -e^{-\beta\gamma_{\rm{eq}}\Delta E_{\rm{p}}}}~,
\label{eq:knock6}
\end{eqnarray}
where Eq.~(\ref{eq:knock6}) is obtained from Eq.~(\ref{eq:knock5}) by substituting Eq.~(\ref{eq:phiofE}). Hence the stationary distribution $g_{\rm eq}(E_{\rm{p}})=g(E_{\rm{p}},\infty)$ takes the form 
\begin{eqnarray}  
g_{\rm eq}(E_{\rm{p}})  & = & \frac{\omega_{\rm eq}e^{\beta\gamma_{\rm{eq}} E_{\rm{p}}}}{N_{\rm{v}}\Gamma_0} \phi(E_{\rm{p}})
\label{eq:knock7}\\
 & = &\frac{\beta\gamma_{\rm{eq}} e^{\beta\gamma_{\rm{eq}} E_{\rm{p}}}}{e^{\beta\gamma_{\rm{eq}}\Delta E_{\rm{p}}} -e^{-\beta\gamma_{\rm{eq}}\Delta E_{\rm{p}}}}~.
\label{eq:knock8}
\end{eqnarray}
Note that the stationary state always exists provided that the integral in Eq.~(\ref{eq:knock5}) is finite.

In practice, $g(E_{\rm{p}},t)$ is inherently time dependent, because $\gamma$ is time dependent.  Figure~\ref{fig:Ehist} graphs $g(E_{\rm{p}},t)$ from the Monte-Carlo simulations discussed in the previous section ($\Delta E_{\rm{p}}/E_0=0.75$) at five equally spaced times (\emph{solid, light grey} curves).  We construct the stationary $g_{\rm{eq}}(E_{\rm{p}})$ (\emph{solid black} curve), using $\gamma = \langle\gamma\rangle$ taken from $p(\gamma)$ (\emph{bottom right} panel of Fig.~\ref{fig:basicdE}), by averaging $g(E_{\rm{p}})$ from $50$ snapshots in time.  For comparison, we also plot $g_{\rm{eq}}(E_{\rm{p}})$ from Eq.~(\ref{eq:knock8}) (\emph{dashed} curve).  The \emph{light grey} curves display the instantaneous $g(E_{\rm{p}},t)$, at five equally spaced times. For both the average and instantaneous $g_{\rm{eq}}(E_{\rm{p}})$, there is a systematic overpopulation of strong pinning sites compared to weak pinning sites, illustrating the excavation property illustrated in Fig.~\ref{fig:Ehist_diag}.  We attribute the discrepancy between $g(E_{\rm{p}})$ and Eq.~(\ref{eq:knock8}) to the fact that Eq.~(\ref{eq:masterg}) does not allow for $\gamma$ to vary in response to glitches.  We also note that, in results not presented here, the simulated $g_{\rm{eq}}(E_{\rm{p}})$ is better approximated by Eq.~(\ref{eq:knock8}) when synchronous automaton rules are used.

\begin{figure*} 
\includegraphics[scale=0.36,angle=90]{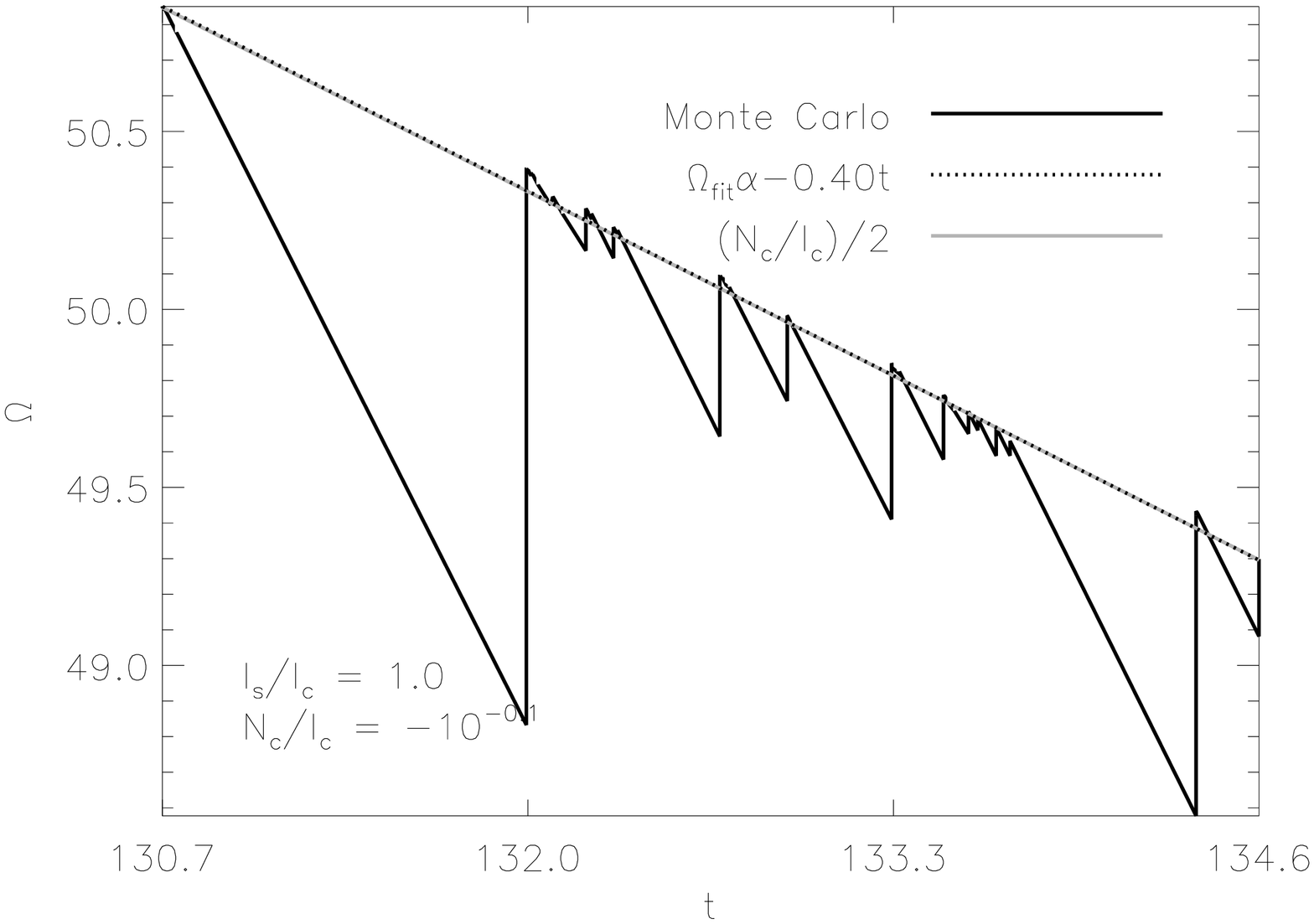}
\includegraphics[scale=0.36,angle=90]{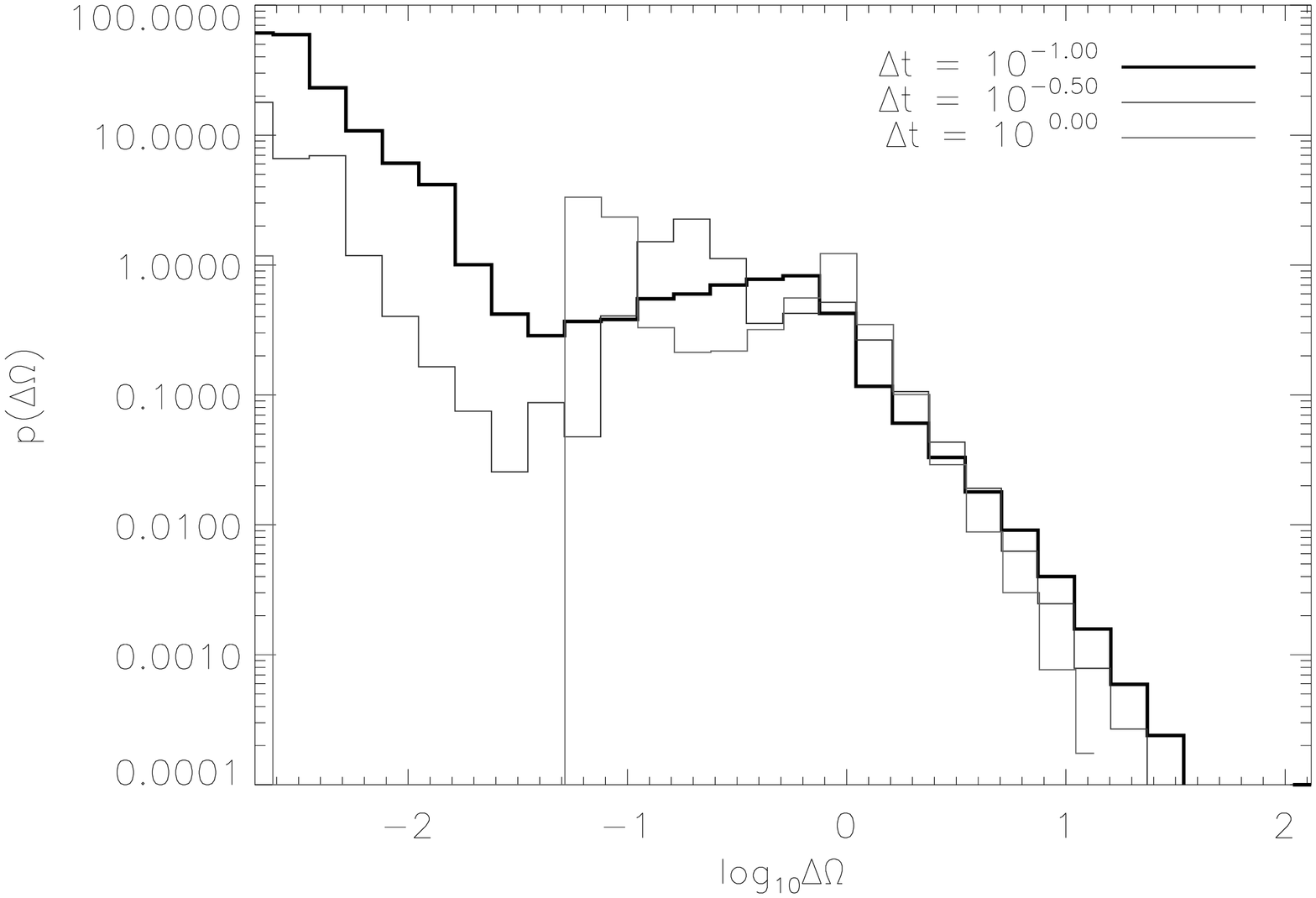}
\includegraphics[scale=0.36,angle=90]{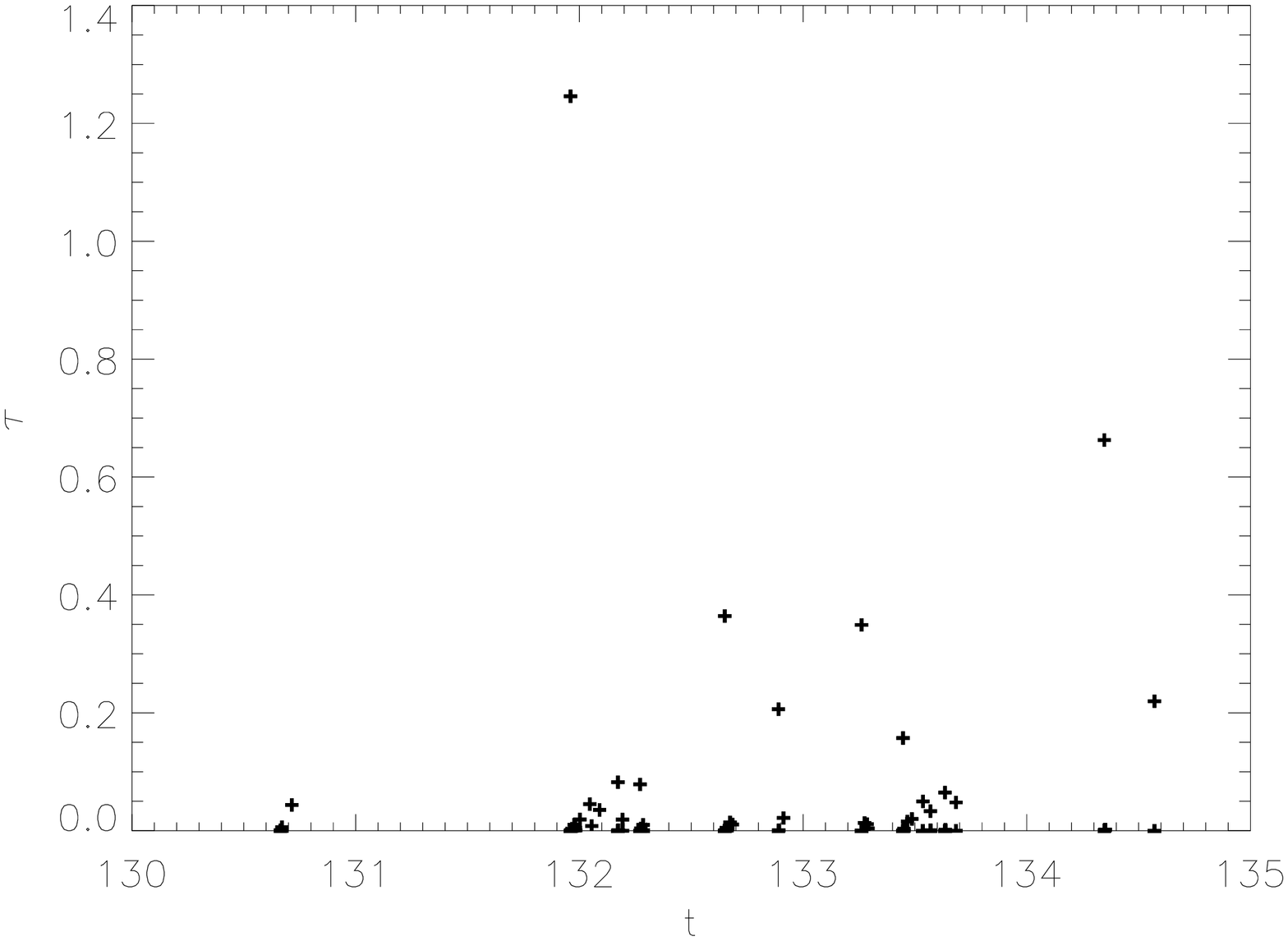}
\includegraphics[scale=0.36,angle=90]{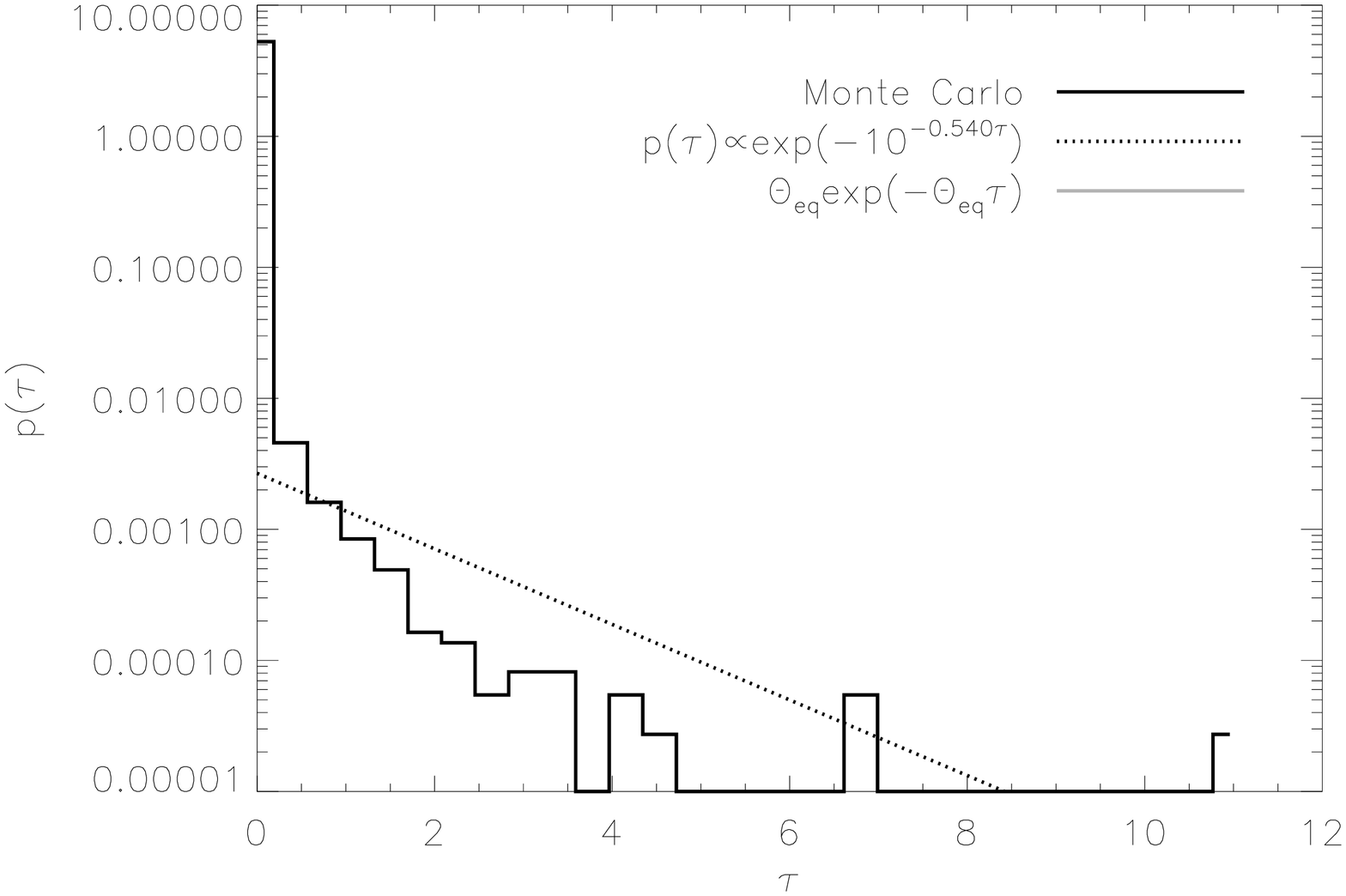}
\includegraphics[scale=0.36,angle=90]{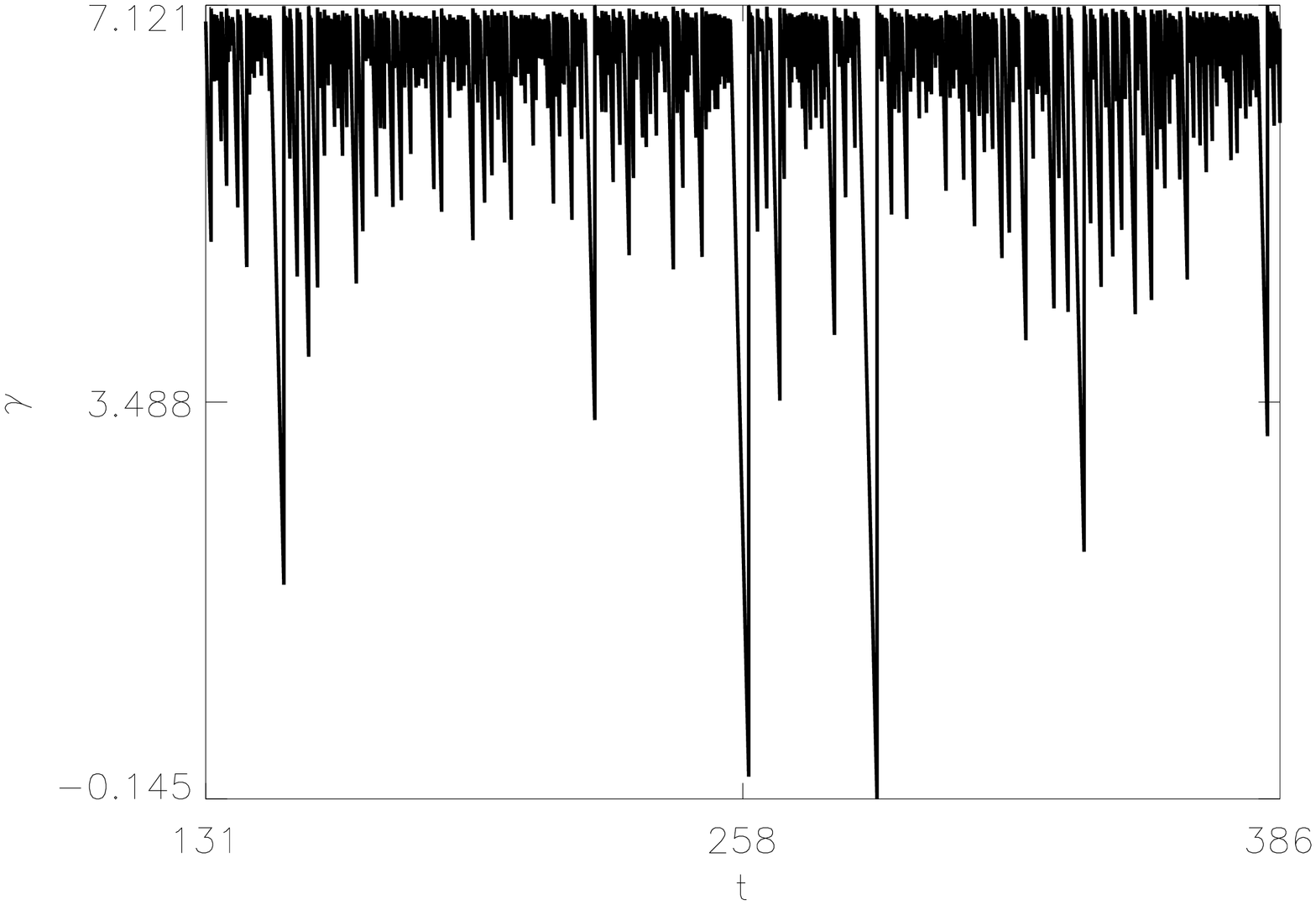}
\includegraphics[scale=0.36,angle=90]{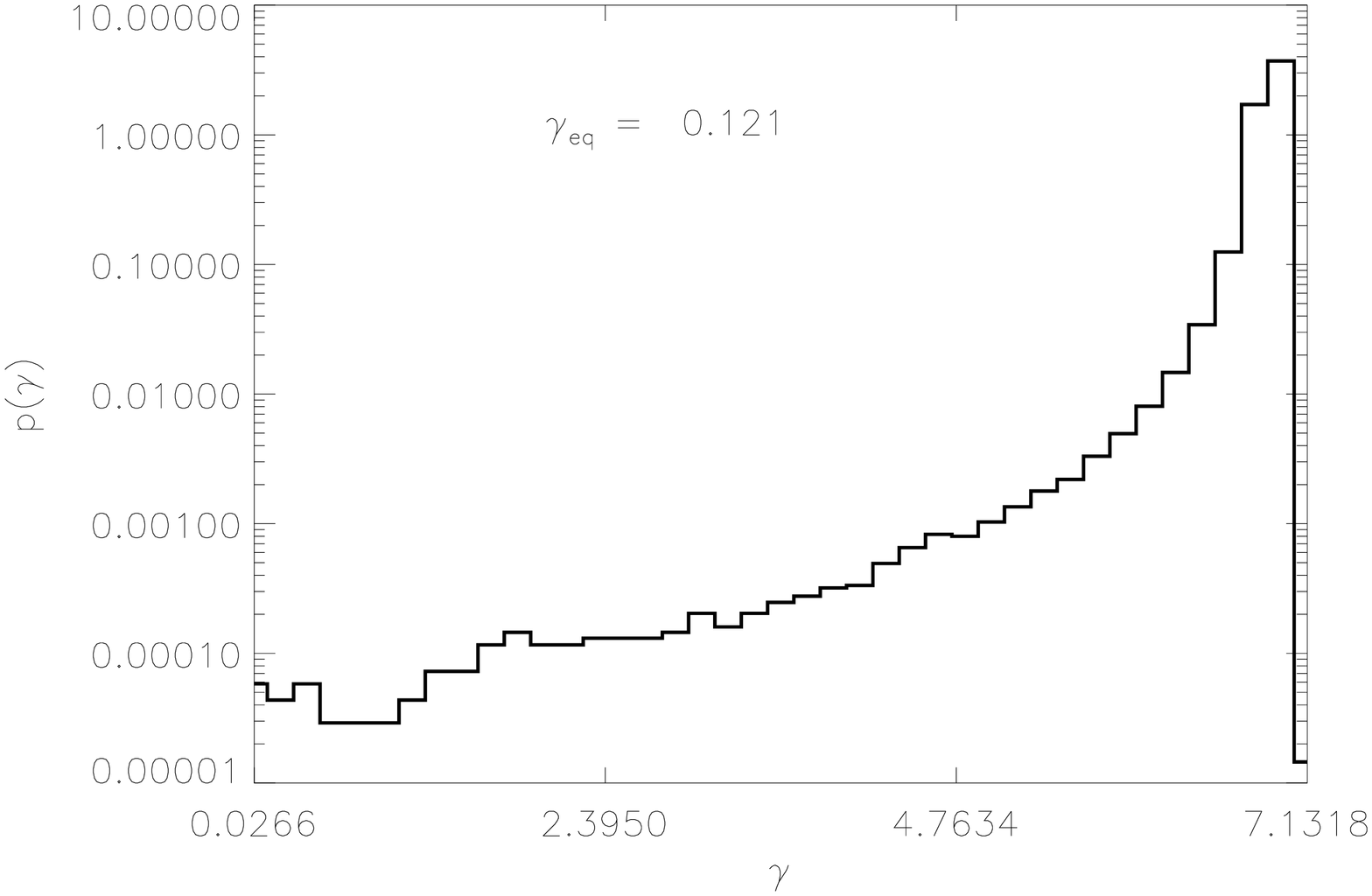}
\caption{Results from Monte-Carlo simulations of a system with asynchronous update rules and knock-on between vortices.  Results are reported as time series and probability density functions (PDFs).  \emph{Top left}:  Angular velocity as a function of time, $\Omega_{\rm{c}}(t)$, for the time interval $130.7<t<134.6$ (\emph{solid} curve).  The \emph{dotted} curve is an empirical linear fit to $\Omega_{\rm{c}}(t)$ with slope $-0.40$.  \emph{Top right}:  PDF of number of vortices that unpin in each avalanche (\emph{solid black} curve); a power-law fit with power-law index $-1.98$ is overplotted (\emph{dotted} curve).   \emph{Bottom left}:  Time series of $\gamma$.  \emph{Bottom right}:  PDF of $\gamma$.  Simulation parameters:  $N_{\rm{v}}=500$ vortices with pinning energies in the range $E_{\rm{p}}\in[E_0-\Delta E_{\rm{p}},E_0+\Delta E_{\rm{p}}]$, $N_{\rm{c}}/I_{\rm{c}}=10^{-0.1}$, $\alpha=10^{-4}$, $I_{\rm{s}}/I_{\rm{c}}=1.0$, $E_0=20.0$, $\Delta E_{\rm{p}}/E_0=1$ and $N_t=5\times 10^5$. }
\label{fig:avalanche}
\end{figure*}

\begin{figure*}
\includegraphics[scale=0.3,angle=90]{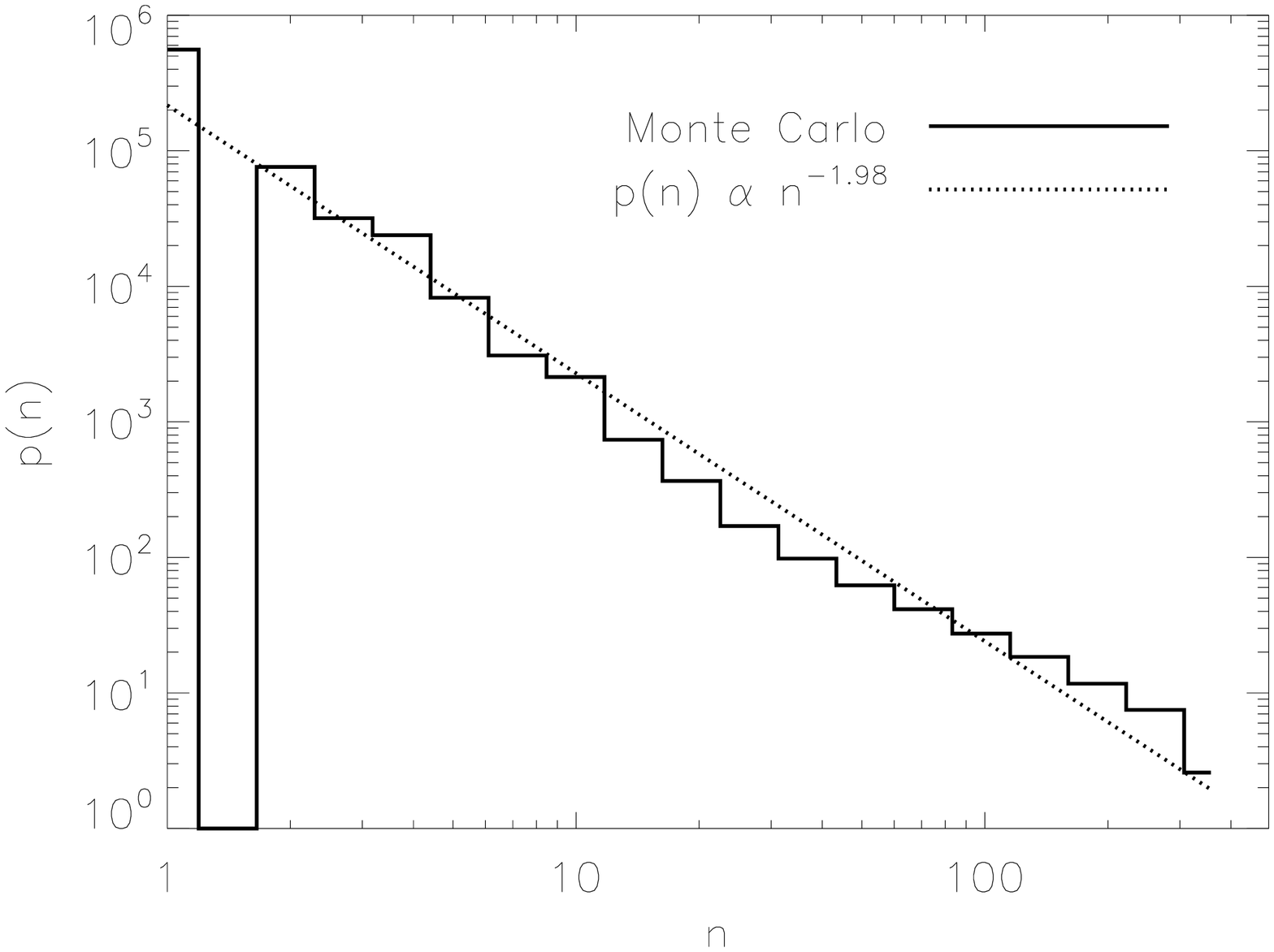}
\includegraphics[scale=0.3,angle=90]{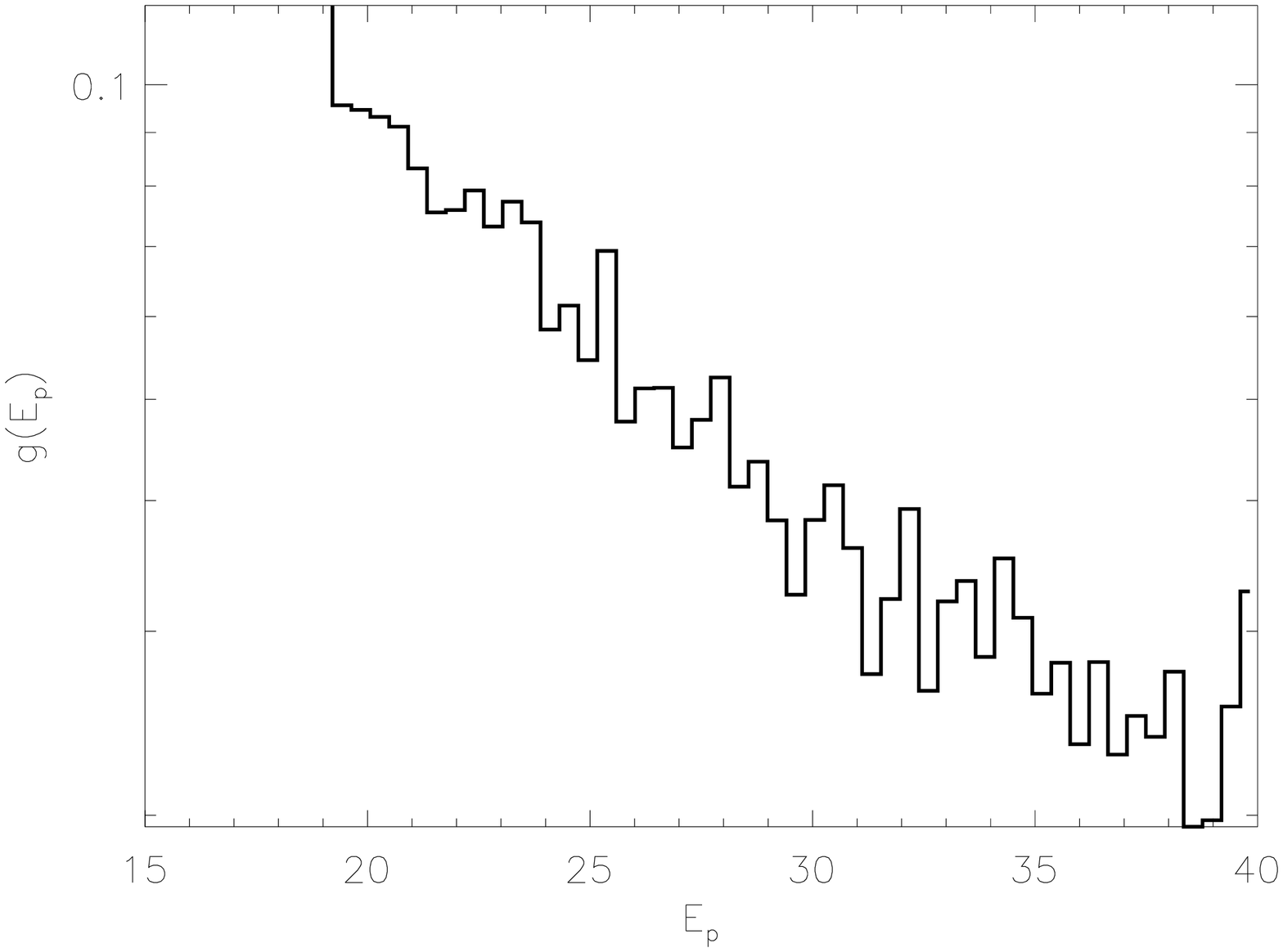}
\caption{\emph{Left}: PDF of number of vortices that unpin in each avalanche, $n$, for the same simulations as described in Fig.~\ref{fig:avalanche}.  This should be compared to the \emph{top right} panel of Fig.~\ref{fig:avalanche}, where glitch size is given as $\Delta\Omega$.  \emph{Right}:  $g(E_{\rm{p}})$ for simulations reported in Fig.~\ref{fig:avalanche} (\emph{solid} histogram).  The bias towards weak pinning sites contradicts the excavation concept described in Sec.~\ref{subsec:excavate}.}
\label{fig:avalanchedE}
\end{figure*}

\begin{figure*}
\begin{center}
\includegraphics[scale=0.325,angle=90]{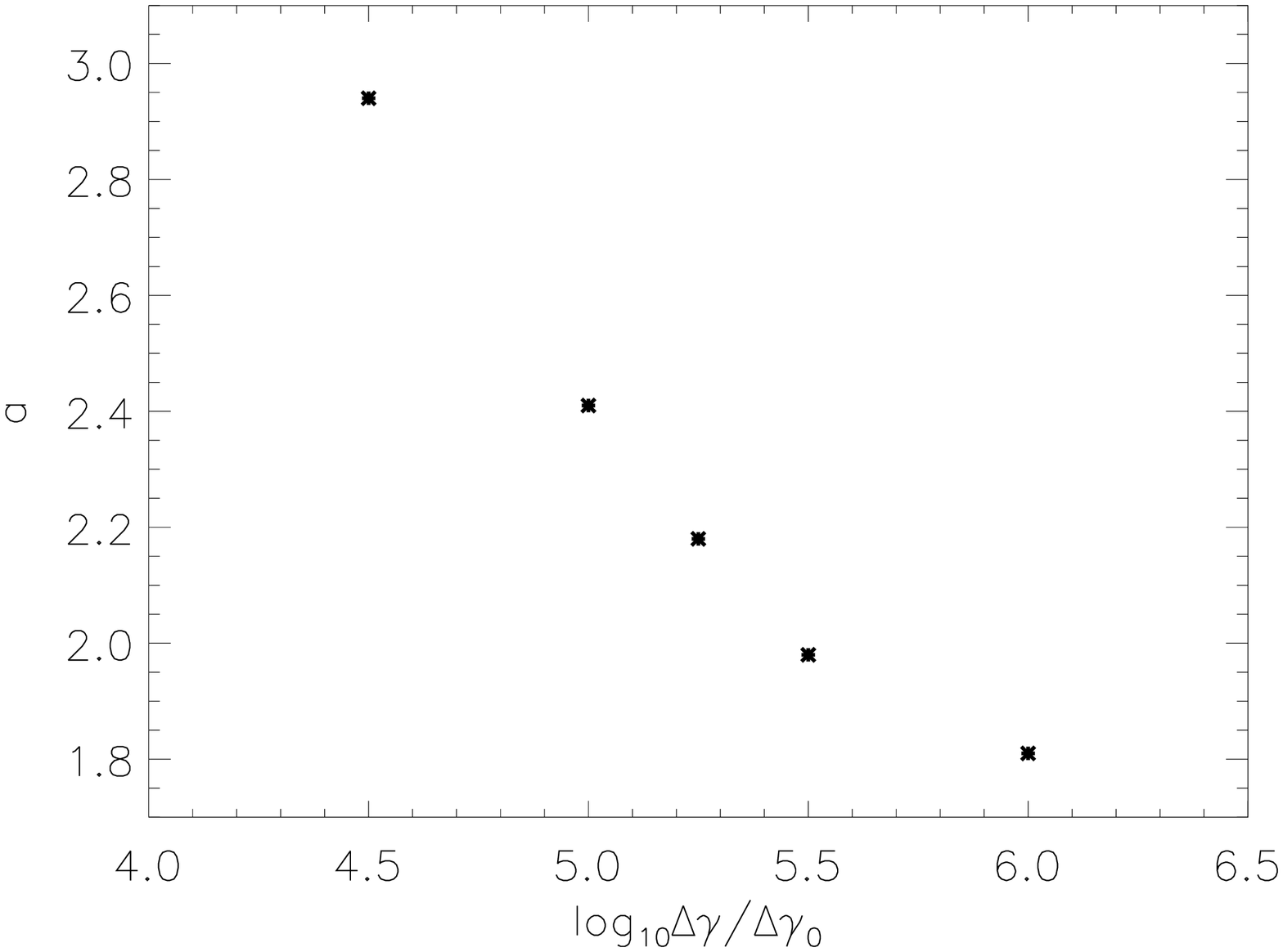}
\end{center}
\caption{Best-fit power-law indices for the glitch size (measured in number of vortices unpinned) distribution resulting from Monte Carlo simulations which include knock on between unpinned and still-pinned vortices.  Knock on is parametrised by a boost in the unpinning rate due to unpinned vortices moving through the pinned vortex lattice.  The PDF is constructed as a function of this boost, $\Delta\gamma/\Delta\gamma_0$ for Monte Carlo simulations; $\Delta\gamma_0$ is given in Eq.~(\ref{eq:dgam0}).  Simulation parameters:  $N_{\rm{v}}=500$ vortices with pinning energies in the range $E_{\rm{p}}\in[E_0-\Delta E_{\rm{p}},E_0+\Delta E_{\rm{p}}]$, $N_{\rm{c}}/I_{\rm{c}}=10^{-0.1}$, $\alpha=10^{-4}$, $I_{\rm{s}}/I_{\rm{c}}=1.0$, $E_0=20.0$, $\Delta E_{\rm{p}}/E_0=1$ and $N_t=5\times 10^5$. }
\label{fig:powerlaw}
\end{figure*}
\section{Vortex avalanches}\label{sec:branching}

In Sec.~\ref{sec:MCsimple}-\ref{sec:heteropinning}, we find that when each vortex unpins according to an independent Poisson process and inter-vortex correlations are ignored, vortex unpinning avalanches with a broad range of sizes do not occur, whether or not multiple pinning energies are available.  Indeed, vortices unpin one by one and the only way to get events that involve $n>1$ vortices is to sum them together over an observational window $\Delta t$ (see Sec.~\ref{subsec:glitchfind}).  Therefore, to explain the catastrophic, near-instantaneous unpinning of vortices in groups that is responsible for pulsar glitches, collective unpinning mechanisms must be considered.  In this section we quantify how unpinning events multiply when a recently unpinned vortex triggers unpinning of another pinned vortex, laying the basis for a domino-like knock-on process.  Results from first-principles, quantum mechanical simulations of single vortex unpinning and repinning \citep{Warszawski:2010individual} motivate this study.  They point to two knock-on mechanisms:  unpinning by the Magnus-like proximity effect, and unpinning by acoustic waves.  In what follows, we focus on the former.

According to the Feynman-Onsager quantisation condition, the change in the superfluid speed at one of the vortices, $\Delta v$, when the separation between a moving vortex and its nearest neighbour decreases from $d_{\rm{F}}$ to $\epsilon d_{\rm{F}}$ ($\epsilon <1$) is given by \citep{CHENG:1988p180}
\begin{eqnarray}
\label{eq:deltavknock}
 \Delta v&=&\frac{\kappa}{2\pi \epsilon d_{\rm{F}}}\frac{1-\epsilon}{\epsilon}~.
\end{eqnarray}
Just as in the global shear lowers the pinning barrier via $\gamma$, Eqs.~(\ref{eq:deltavknock}) and (\ref{eq:Magnus}) demonstrate that a local over-density of vortices (relative to the unstressed lattice configuration) lowers the pinning barrier too. Locally the shear parameter $\gamma$ changes from $\gamma$ to $\gamma-\Delta\gamma_0$, with
\begin{eqnarray}
\label{eq:dgam0}
\Delta\gamma_0 &=& \frac{\kappa}{2\pi d_{\rm{F}}R\Delta\Omega_{\rm{cr}}}\frac{1-\epsilon}{\epsilon }~.
\end{eqnarray}

Suppose, for the sake of illustration, that all the vortices are evenly spaced by $d_{\rm{F}}$ and one vortex unpins; call this a \emph{trigger event}.  The unpinned vortex moves radially, approaches a nearby vortex, and reduces the latter's pinning barrier.  The time-scale for this to occur is $\tau_{\rm{k}}=d_{\rm{F}}/v_{\rm{v}}\sim10^{-11}\,\rm{s}$, where $v_{\rm{v}}= R\Delta\Omega$ is the radial component of the unpinned vortex velocity.  During this time interval the unpinning threshold is reduced by $\Delta\gamma$, such that the unpinning rate for a single vortex becomes 
\begin{equation}
\label{eq:knockrate}
 \lambda(E_{\rm{p}})=\Gamma_0 e^{-\beta E_{\rm{p}}(\gamma-\Delta\gamma)}~.
\end{equation}
So, each trigger unpins, on average, $\lambda(E_{\rm{p}})\tau_{\rm{k}}$ vortices. In practice, to see whether or not this formulation of knock on results in unpinning avalanches, Eq.~(\ref{eq:knockrate}) should be incorporated into the automaton rules described in Sec.~\ref{subsec:asynch}, which we do in the next section.

\subsection{Automaton rules}\label{subsec:avaauto}

In this section, we modify the automaton rules described in Sec.~\ref{sec:MCsimple} to include knock-on.
\begin{enumerate}
 \item For each vortex $i$, a time $\tau_i$ until the next unpinning is drawn from $p(\tau_i,E_{\rm{p}},t)$, with the initial condition  $\gamma(t=0)=\gamma_0$.
 \item The system is advanced in time by the minimum waiting time $\tau_{\rm{min}}=\rm{min}\{\tau_i\}$, $t_n=t_{n-1}+\tau_{\rm{min}}$.
 \item Increment $\Omega_c$ by $\Delta\Omega_c=\tau_{\rm{min}}N_{\rm{c}}/I_{\rm{c}}$.  The state parameter $\gamma(t+\tau_{\rm{min}})$ becomes $\gamma(t+\tau_{\rm{min}})=\gamma(t)+(\Delta\Omega_c-\Delta\Omega_s)/\Delta\Omega_{\rm{cr}}$.  Now allow the avalanche to proceed.
 \begin{enumerate}
  \item Ask each vortex (other than the original trigger) if it unpins, with probability $1-\exp (-\lambda\tau_{\rm{k}})$ [the probability of one or more events during an interval $\tau_{\rm{k}}$, characteristic of the knock-on process, in a Poisson process].  $\Delta \gamma$ is parametrised in terms of $\Delta\gamma_0$ [Eq.~(\ref{eq:dgam0})]. 
  \item If at least one extra vortex unpins, ask all the still-pinned vortices if they unpin until none do and the avalanche is over.  The avalanche size is proportional to the total number of vortices that have unpinned, $n$.
 \end{enumerate}
 \item\label{step3} Decrement $\Omega_s$ by $\Delta\Omega_s=n\Delta L/I_{\rm{s}}$, so that $\Omega_c$ changes by $\Delta\Omega_c=-n\Delta L/I_{\rm{s}}$.  The shear parameter $\gamma(t+\tau_{\rm{min}})$ is updated to $\gamma(t+\tau_{\rm{min}})=\gamma(t)+(\Delta\Omega_c-\Delta\Omega_s)/\Delta\Omega_{\rm{cr}}$.
 \item All unpinned vortices are assigned new waiting times (and new pinning strengths if $\Delta E_{\rm{p}}\neq 0$), drawn from $p(\tau_i,E_{\rm{p}},t+\tau_{\rm{min}})$.  Note that this is the \textit{second} $\gamma$ update within a single time step.
 \item For all vortices that did not unpin in the previous interval, the time until the next unpinning event for the $i^{\rm{th}}$ vortex at the $n^{\rm{th}}$ iteration, $\tau_{i,n}$ becomes \begin{equation}                                                                                                                                  
     (\tau_{i}-\tau_{\rm{min}})\exp\left[-\beta E_0 \gamma(t+\tau_{\rm{min}})\right]/\exp\left[-\beta E_0 \gamma(t)\right]~,                                                                                                                          \end{equation}
      which stretches the remaining fraction of the waiting time for still-pinned vortices in response to the new state of the system.
 \item Repeat steps 2--5.
\end{enumerate}
The above rules differ from those given in Sec.~\ref{subsec:asynch}.  Each unpinning is treated as a trigger for subsequent unpinnings, which occur at a higher mean rate, causing an unpinning avalanche.    In Sec.~\ref{subsec:asynch}, by contrast, the subsequent unpinnings are not included, such that $n=1$ always.

\subsection{Output statistics}\label{subsec:avamc}

The impact of knock-on on the event size distribution depends on the relative strength of pinning and the knock-on variable, $\Delta\gamma$ [we discuss the physical implications of different $\Delta\gamma$ in Sec.~(\ref{sec:master})].  To understand why, first consider a weak, single-pinning-energy scenario, in which a relatively small $\Delta\Omega$ (high $\gamma_{\rm{eq}}$) is sufficient to unpin vortices.  In this case, $\Delta\gamma$ is small compared to $\gamma_{\rm{eq}}$, and hence the increase in the unpinning rate due to the knock-on effect is small.  On the other hand, for strong pinning, $\gamma_{\rm{eq}}$ is smaller, rendering knock-on more effective.  In this way, either increasing $E_{\rm{p}}$ or decreasing $\Delta\gamma$ drives the system towards avalanches.  

Results from Monte-Carlo simulations that employ the asynchronous update rules described above (Sec.~\ref{subsec:avaauto}), are presented in Fig.~\ref{fig:avalanche}.  In the \emph{top left}, we graph $\Omega_{\rm{c}}(t)$ (\emph{solid black} curve), overplotted with a linear fit (\emph{dotted black} curve), as well as the curve that corresponds to uniform spin down [slope $-N_{\rm{c}}/(2I_{\rm{c}})$, \emph{solid grey} curve].  Unlike the corresponding plots in Fig.~\ref{fig:basic} and \ref{fig:basicdE}, $\Omega_{\rm{c}}(t)$ is punctuated by abrupt jumps, caused by near-instantaneous knock-on events (the characteristic knock-on time scale, quantified in Sec.~\ref{subsec:knockbranch}, satisfies $\tau_{\rm{k}}\ll \langle\tau\rangle$), followed by periods of uniform deceleration.    On long time scales, the spin down matches that expected for a container filled with an unpinned, smoothly decelerating superfluid [Eq.~(\ref{eq:feedback}) with $d\Omega_{\rm{c}}/dt=d\Omega_{\rm{s}}/dt$ and $I_{\rm{s}}/I_{\rm{c}}=1$, gives $d\Omega_{\rm{c}}/dt = N_{\rm{c}}/(2I_{\rm{c}})$], as shown by the agreement between the fitted (\emph{dot-dashed}) and linear (\emph{grey}) spin-down curves. 

The \emph{top right} panel of Fig.~\ref{fig:avalanche} graphs $p(\Delta\Omega_{\rm{c}})$, calculated using three $\Delta t$ (Sec.~\ref{subsec:glitchfind}).  Interestingly, $p(\Delta\Omega_{\rm{c}})$ spans over three decades in $\Delta\Omega_{\rm{c}}$, as compared to the narrow $\Delta\Omega_{\rm{c}}$ range without knock-on (see Fig.~\ref{fig:basic} and \ref{fig:basicdE}).  There is a plateau between $0.1\lesssim\Delta\Omega_{\rm{c}}\lesssim 1.0$ for the two smaller values of $\Delta t$.  Further investigation is required to understand this feature.   

We also measure glitch size by the number of vortices that unpin during an avalanche, $n$.  Quantitative agreement between this approach and the glitch-finding algorithm in Sec.~\ref{subsec:glitchfind} is best for $\theta\Delta t\ll 1$when only one event is captured in $\Delta t$.  The \emph{left} panel of Fig.~\ref{fig:avalanchedE} graphs the PDF of $n$ (\emph{solid black} curve); a power-law fit is graphed as a \emph{dotted} curve. In this example, with $\lambda\tau_{\rm{k}}\sim 1$ ($\lambda$ is a function of $\gamma$, and hence $t$), $p(n)$ covers two decades.  In simulations not shown here, we investigated how $p(n)$ changes with the strength of the knock-on parameter, whilst keeping $E_{\rm{p}}$ constant.  We find that for $\Delta\gamma/\Delta\gamma_0<10^3$, $p(n)$ is sharply peaked (as in the \emph{top right} panels of Fig.~\ref{fig:basic} and \ref{fig:basicdE}).  For $10^3<\Delta\gamma/\Delta\gamma_0<10^6$, $p(n)$ extends over several decades, and is well described by a power law, as in the example plotted in Fig.~\ref{fig:avalanche} and \ref{fig:avalanchedE}.   For $\Delta\gamma/\Delta\gamma_0>10^6$, $\lambda\tau_{\rm{k}}$ exceeds unity, and hence every trigger event results in a system-spanning avalanche (\emph{ie}. all vortices unpin).  In this regime, the power law changes into a hump at large $n$. 

The fact that the system is sensitive to $\lambda\tau_{\rm{k}}$, rather than self-organising to always produce power-law-distributed glitches, reminds us that the primary trigger events can decelerate the superfluid quickly enough to keep pace with the crust; avalanches are not essential and only occur when $\Delta\gamma$ is large.  As demonstrated in  Fig.~\ref{fig:avalanche}, however, if  $\Delta\gamma$ is too large, system-spanning avalanches are inevitable as soon as a trigger event occurs, leading to the overpopulation of large glitch events.  This behaviour is quantified in Sec. \ref{subsec:knockbranch}.  Figure~\ref{fig:powerlaw} shows that the power-law index, $a$, decreases for increasing $\Delta\gamma$, indicating that larger glitches become more probable as the knock-on mechanism strengthens.

Waiting times as a function of glitch epoch and the waiting-time PDF are graphed in the \emph{centre} panels of Fig.~\ref{fig:avalanche}.  The PDF spikes at small $\tau$, and there is significant statistical weight in the tail, extending out to $\tau = 11$.  Excising the initial spike, and considering only $\tau<4$, we fit an exponential (\emph{dotted} curve)  to the entire data set.  $p(\tau)=\theta(\gamma_{\rm{eq}})\exp[-\theta(\gamma_{\rm{eq}})\tau]$ is not a good fit to the data.  To understand this, we consider the PDF of $\gamma$ in the \emph{bottom right} panel of Fig.~\ref{fig:avalanche}.  We find $\gamma_{\rm{eq}}=0.121\ll 6.95=\langle\gamma\rangle$, so $\theta_{\rm{eq}}$ significantly overestimates the unpinning rate.  This is not surprising since Eq.~(\ref{eq:eqgam}) defines a lag that ensures that the \emph{trigger} unpinning rate is enough to decelerate the superfluid with the crust.  Large jumps in the time series of $\gamma$ ($\Delta\gamma/\gamma\sim 1$, \emph{bottom left} panel), corresponding to abrupt $\Omega_{\rm{c}}$ jumps seen in the \emph{top left} panel, explain why $p(\gamma)$ is broad (a range of 7.11 compared to 0.4 in the simple, no-knock-on case). 

The time-averaged PDF of occupied pinning energies, $g(E_{\rm{p}})$, is graphed in Fig.~\ref{fig:avalanchedE}.  The bias towards weak pinning energies is in stark contrast to the no-knock-on case described in Fig.~\ref{fig:Ehist} in which $g(E_{\rm{p}})$ is excavated.  This is because the scale invariance promoted by $\gamma$ variability outweighs the escavation effect.

\subsection{Knock-on as a branching process}\label{subsec:knockbranch}

From the Monte-Carlo simulations, we conclude that knock-on is effective in catalysing avalanche events.  However, a power-law event size distribution only arises when the strength of the knock-on mechanism is tuned so that the boosted unpinning rate after an initial trigger is just sufficient to raise the probability of a subsequent unpinning to unity.  This is the condition $\lambda\tau_{\rm{k}}=1$ referred to above.  In this section we make this result precise using the theoretical framework of branching processes.

From Sec.~\ref{sec:branching}, we know that each trigger unpins, on average, $\lambda(E_{\rm{p}})\tau_{\rm{k}}$ vortices.  Note that $\tau_{\rm{k}}$ is the time-scale for a single \emph{branch} in the avalanche, \emph{not} the duration of the entire avalanche.  The number of branches is typically the logarithm of the total number of vortices that unpin in the avalanche, and hence the avalanche typically lasts for $\lesssim 10^2\tau_{\rm{k}}$. 

\begin{center}
\begin{table}
\begin{minipage}[b]{0.5\linewidth}\centering
\begin{tabular}{c l l}
 \hline Quantity			& Typical pulsar 		& Units \\\hline
 		$R$ 				& $10^4$				& m \\
 		$\Omega_{\rm{c}}$ 	& $2\pi\times 10^2$		&$\rm{rad\,s}^{-1}$\\
 		$N_{\rm{c}}/I_{\rm{c}}$ 		& $10^{-13}$			&$\rm{rad\,s}^{-2}$ \\
 		$\rho$ 				& $10^{17}$				&$\rm{kg\,m\,s}^{-1}$\\
 		$\kappa$ 			& $10^{-7}$				&$\rm{m}^2\rm{s}^{-1}$\\
 		$\Gamma_0$ 			& $10^{22}$				&$\rm{s}^{-1}$\\
 		$E_{\rm{p}}$ 		& $1$			&$\rm{MeV}$ \\
 		$d_{\rm{F}}$ 		& $10^{-5}$			&$\rm{m}$ \\
 		$\Delta\Omega_{\rm{cr}}$ &	$10^{-3}$			&$\rm{rad\,s}^{-1}$\\
 		$I_{\rm{s}}/I_{\rm{c}}$ & $10^{-2}$			&---\\
 		$N_{\rm{v}}$ 		&  $10^{18}$			&---\\
 		$\alpha$ 			& $10^{-9}$			&---\\\hline
\end{tabular}
\end{minipage}
\caption{Fiducial pulsar parameters.}
\label{tab:paramknock}
\end{table}
\end{center}

The trigger and subsequent knock-on events form the basis of a Galton-Watson-Bienaym\'{e} branching process, which generates events that follow a generalised Poisson distribution \citep{Dobson:2007p8221}.  The fate of familial lineages originally spurred the study of branching processes \citep{Harris}.  More recently, branching processes have been used to describe natural phenomena such as cascades of splitting particles and polymerisation \citep{Tobita:1989}, statistical studies of DNA \citep{Avise:1984}, the formation of random fractal sets \citep{Jagers:2005}, and models of electricity grids \citep{Carreras:2002p8220}.  

Consider a composite branching process in which the triggers follow a Poisson distribution, with mean number of triggers $\theta\tau_{\rm{trig}}$ in time $\tau_{\rm{trig}}$.  Each trigger and subsequent knock-on unpinning causes, on average, a further $\lambda\tau_{\rm{k}}$ ($\leq 1$) vortices to unpin.  The event size distribution, $p(n;\tau_{\rm{trig}})$, for the composite process is given by \citep{Consul:1973,Dobson:2007p8221}
\begin{equation}
\label{eq:pofn}
 p(n;\tau_{\rm{trig}}) = \theta\tau_{\rm{trig}} \left(n\lambda\tau_{\rm{k}}+\theta\tau_{\rm{trig}}\right)^{n-1}\frac{e^{-n\lambda\tau_{\rm{k}}-\theta\tau_{\rm{trig}}}}{n!}\,,
\end{equation}
where $n$ is the total number of vortices that unpin in a single avalanche.  As discussed in Sec.~\ref{sec:MCsimple}, after each avalanche, the lag between the crust and superfluid adjusts, reducing the unpinning rate everywhere.  Therefore, there is no unique waiting time for the trigger events, as the rate varies on the same time-scale that the triggers recur.  Equation~(\ref{eq:pofn}) then implies a
mean event size and mean unpinning rate if one waits for a time $\theta^{-1}$ (which picks up one trigger on average) of
\begin{eqnarray}
 \langle n\rangle&=&\frac{1}{1-\lambda\tau_{\rm{k}}}~,
\end{eqnarray}
and
\begin{eqnarray}
 \frac{d\langle n\rangle}{dt}=\frac{\theta}{1-\lambda\tau_{\rm{k}}}~
\end{eqnarray}
respectively.

The shape of $p(n)$ depends strongly on $\lambda$.  For $\lambda\tau_{\rm{k}}\approx 1$, which corresponds to each unpinning in an avalanche catalysing one other on average, $p(n)$ exhibits a power-law tail with exponent $-3/2$. To see this mathematically, we use Stirling's approximation $n!\approx \sqrt{2\pi n}\left(n/e\right)^n$, to obtain
\begin{eqnarray}
 \lim_{n \to \infty} p(n)&\approx& \lambda^{n-1}\left(1+\frac{1}{n\lambda\tau_{\rm{k}}}\right)^{n-1}\frac{e^{n-n\lambda\tau_{\rm{k}}}}{n\sqrt{2\pi n}}~,\\
 \lim_{n \to \infty, \lambda\tau_{\rm{k}} \to 1} p(n) &\approx& \frac{n^{-3/2}}{\sqrt{2\pi }}~.
\end{eqnarray}

For a state-independent system, the condition $\lambda\tau_{\rm{k}}=1$ leading to a power-law size distribution requires a high level of \textit{fine tuning}.  Even with the self-organising influence of torque feedback, fine tuning, in the sense of $\lambda\tau_{\rm{k}}=1$, does not arise naturally.  Torque feedback ensures $d\langle n\rangle/dt$ is the correct value to accommodate the electromagnetic spin-down torque imposed externally, but it does nothing to push $\lambda\tau_{\rm{k}}$ towards unity.  In other words, vortex unpinning in non-power-law-distributed avalanches is sufficient to decelerate the superfluid at the same rate as the crust on average (see the \emph{top left} panel of Fig.~\ref{fig:basic}, for example), even without $\lambda\tau_{\rm{k}}=1$.  

The fine tuning condition can be re-expressed as
\begin{equation}
\label{eq:finetune}
\beta E_{\rm{p}} =  -(\gamma_{\rm{eq}}-\Delta\gamma)^{-1}\ln \Gamma_0.
\end{equation}
Equation~(\ref{eq:finetune}) can variously be interpreted as a condition on pinning strength or stellar temperature, neither of which are regulated by vortex unpinning.  That is, $\beta$ is determined independently by stellar cooling, $\gamma_{\rm{eq}}$ results from the mean torque, $\Delta\gamma$ is a direct result of quantum mechanics and the pinning landscape, and $\Gamma_0$ is also quantum mechanical.  For a typical pulsar, with physical parameters given in Table~\ref{tab:paramknock} and $\epsilon = 1/2$, we obtain $\Delta\gamma_0\sim 10^{-4}$ and $\lambda\ll 1$, which does not satisfy the stated condition for an avalanching system.  

An alternative collective mechanism is the accumulation of sound waves from many unpinning events, which vibrate the superfluid and increase the unpinning rate by raising the effective temperature, $\beta^{-1}$.  The estimate of the energy emitted in acoustic radiation by a vortex ($\Delta E_{\rm{sound}}=10^{-12}~\rm{J\,m}^{-1}$) provided at the end of \cite{Warszawski:2010individual} suggests that single unpinning events do not output sufficient acoustic energy to noticeably change the system temperature.  However, the cumulative effect of many vortices unpinning may be enough to significantly alter the unpinning rate; of course, we are still obliged to identify a large-scale unpinning trigger.  We defer this investigation to further work.

\section{Future improvements}\label{sec:concknock}

The framework presented in this paper can be extended in many
profitable directions, two of which are outlined briefly here.

\subsection{Towards an analytic description: master equation}\label{sec:master}

In a state-dependent system, we are interested in the probability of occupying a given state.  The relevant state variable in the superfluid-crust system is the differential rotation, parametrised by $\gamma$.  Following \cite{Daly:2007p8447}, we employ the Chapman-Kolmogorov forward equation to track the evolution of the probability $p(\gamma,t)d\gamma$ of being in a state $\gamma$ at time $t$:
\begin{equation}
\label{eq:CK}
  \frac{\partial p(\gamma,t)}{\partial t} = \frac{\partial}{\partial \gamma}\left[ \dot{\gamma}_0 p(\gamma,t) \right]-\Theta(\gamma,t)p(\gamma,t)+
      \int_0^{\infty}d(\Delta\gamma)\Theta(\gamma-\Delta\gamma,t)p(\gamma-\Delta\gamma,t)h(\Delta\gamma,\gamma-\Delta\gamma)~.
\end{equation}
The first term on the right-hand side of Eq.~(\ref{eq:CK}) describes deterministic spin down in response to the electromagnetic torque; we have $\dot{\gamma}_0=N_{\rm{c}}/(I_{\rm{c}}\Delta\Omega_{\rm{cr}})$. The second term is the rate at which the system jumps \emph{away} from the state $\gamma$; $\Theta(\gamma,t)$ is the unpinning rate at a given $\gamma$.  The third term accounts for the rate at which the system jumps \emph{into} the state $\gamma$ via a jump of size $\Delta\gamma$; here, $h(\Delta\gamma,\gamma-\Delta\gamma)$ is the probability density of an event of size $\Delta\gamma$, when the system is in a state $\gamma-\Delta\gamma$.

For a vortex pinned with energy $E_{\rm{p}}$, the unpinning rate is $\Theta(\gamma)=\Gamma_0 e^{-\beta E_{\rm{p}}\gamma}$.  In the simplest case, discussed in Sec.~\ref{sec:feedback}, each event consists of the unpinning, outward motion, and repinning of a single vortex.  There is no knock-on and hence no vortex avalanches.  The jump size distribution in this case is $h(\Delta\gamma,\gamma-\Delta\gamma)=\delta(\Delta\gamma-\Delta\gamma_1)$, where $\Delta\gamma_1$ is the change in $\gamma$ due to the unpinning of a single vortex, and $\delta(\cdot)$ is the Dirac delta function.  More complicated jump distributions can be incorporated into Eq.~(\ref{eq:CK}), via $h(\Delta\gamma,\gamma-\Delta\gamma)$.  For example, rainfall events are modelled by a gamma distribution \citep{Daly:2010}, and hyperexponentials are used in queueing and communication problems \citep{Feldmann:1998}.  Equation~(\ref{eq:pofn}) gives the jump distribution appropriate to a branching process. 

We can calculate the moments of $p(\gamma)$.  Assuming a constant $\Delta\gamma=\Delta\gamma_1$ [\emph{i.e.} $h(\Delta\gamma,\gamma-\Delta\gamma)=\delta(\Delta\gamma-\Delta\gamma_1)$], multiplying Eq.~(\ref{eq:CK}) by $\gamma$, and integrating by parts with respect to $\gamma$, we obtain
\begin{eqnarray}
\label{eq:meangam}
  \frac{d\langle\gamma\rangle}{dt}
    &=&-\dot{\gamma}_0+N\Gamma_0 \Delta\gamma_1\eta(\beta E_{\rm{p}},t)~,
\end{eqnarray}
with $\eta(\beta E_{\rm{p}},t)=\int_0^{\infty}d\gamma e^{-\beta E_{\rm{p}}\gamma}p(\gamma,t)=\langle e^{-\beta E_{\rm{p}}\gamma}\rangle$.  In the steady state, Eq.~(\ref{eq:meangam}) gives
\begin{equation}
\label{eq:eta}
 \eta(\beta E_{\rm{p}})=\frac{\dot{\gamma}_0}{N\Gamma_0\Delta\gamma_1}~.
\end{equation}
Since the right hand side of Eq.~(\ref{eq:eta}) is independent of $\beta E_{\rm{p}}$, $\eta$ must be too. Rearranging Eq.~(\ref{eq:eta}), and assuming $\langle e^{-\beta E_{\rm{p}}\gamma}\rangle\approx  e^{-\beta E_{\rm{p}}\langle\gamma\rangle}$ [a good approximation if $p(\gamma,t)$ is sharply peaked], we arrive at
\begin{equation}
\label{eq:eqgam}
  \langle\gamma\rangle = -\frac{1}{\beta E_{\rm{p}}}\ln\left(\frac{\dot{\gamma}_0}{N\Gamma_0\Delta\gamma_1}\right)=\gamma_{\rm{eq}}~,
\end{equation}
which agrees with Eq.~(\ref{eq:eqgamsimple}).

Solving Eq.~(\ref{eq:CK}) self-consistently is an ongoing avenue of investigation.  We take inspiration from studies of solar flare statistics \citep{Wheatland:2009p8258}, in which a Monte-Carlo algorithm, based on a power-law jump size distribution [$h(\Delta\gamma,\gamma-\Delta\gamma)$], is employed.  Analytic solutions for gamma-function and hyperexponential jump size distributions have also been described by \cite{Daly:2010}.  Heavier tails in $h(\Delta\gamma,\gamma-\Delta\gamma)$ lead to larger excursions from steady spin down, since there are more large events that push the system away from equilibrium.

\subsection{The real pulsar geometry}\label{subsec:realgeometry}

This model does not include the important consideration of the radial position of the vortex, which is relevant to several aspects of the model.  Firstly, the Magnus force experienced per unit vortex length depends directly on velocity through Eq.\ref{eq:Magnus}, with the direct corollary that vortices at different radii experience a different Magnus force, even under the same lag conditions between the superfulid and neutron star crust.  In the model presented here, we make the ansatz that glitch-relevant pinning occurs in only a thin spherical shell in the outer kilometer of the Neutron star, and thus have treated all vortices as having a common radius.  Inclusion of a radial component to such a statistical model would require a corresponding probability distribution of radial position, which would in turn demand an understanding of the radial dependence of pinning strength, as discussed by \cite{Haskell:2011}.  One then expects a fluctuating history of pile-up at different radii. Such spatial inhomogeneity opens up a new channel to drive sandpile-like avalanches \citep{Melatos:2008p204,Warszawski:2008p4510}, supplementing the pile-up that occurs in energy space through the coherent-noise process encoded in $g(E,t)$ \citep{Warszawski:2009p3302}.

The second important effect of radial position is on the length of the vortex and fraction thereof that is intercepted by the crustal pinning region.  The greater the fraction of the vortex submerged in the crust, the greater the energy with which the vortex is pinned, resulting in an increasing function of pinning energy with radial vortex position \citep{Haskell:2011}.  This separate effect would also need to be accounted for in the pinning energy distribution.

It should be noted that the importance of tracking radial position is somewhat lessened if vortices form a turbulent tangle, driven by relative flow between the viscous and inviscid superfluid components, as in the Donnelly-Glaberson instability \citep{Peralta06a,Andersson:2007p196} or its pinning-mediated analogue \citep{Link:2012a,Link:2012b}. In such a scenario, small vortex loops in the tangle behave more like point vortices and the homogeneous statistical model presented  in this paper is a better approximation than otherwise. However, rotation overall polarises the vortex tangle \citep{Mongiovi:2007p214514,Tsubota:2012}, so there is a residual geometric effect, which again would be captured by making $g(E,t)$ depend on cylindrical radius in a more complete model.

\section{Summary}\label{subsec:summaryknock}

In this paper we present the foundations for a statistical model of pulsar glitches.  The model contains three novel features: (1) treatment of vortex unpinning as a Poisson process with variable rate governed by crust-superfluid lag; (2) pinning sites with multiple energies, which lead to excavation of the energy distribution; and (3) knock-on, which catalyses vortex avalanches.  However, the model is incomplete because it requires fine tuning to generate scale-invariant avalanches with a power-law size distribution. 

In Sec.~\ref{sec:feedback}, we demonstrate that torque feedback between the crust and superfluid allows the angular velocity lag between the two components to fluctuate around a constant value.  An average-rate calculation, which assumes that unpinning matches spin down on average, and a more comprehensive description in terms of a  Chapman-Kolmogorov equation (Sec.~\ref{sec:master}), predict the same equilibrium lag, which agrees reasonably well with results from asynchronous Monte Carlo simulations with a single pinning energy.  Importantly, however, torque feedback alone is insufficient to catalyse a broad range of glitch sizes over many decades, even with a broad range of pinning energies.  This result is significant, as it means that correlated vortex unpinning is essential to explain glitch data.

In Sec.~\ref{sec:branching}, we introduce a parametrisation of knock-on between unpinned and still-pinned vortices, which makes it easier for additional vortices to unpin following an initial unpinning. We remain agnostic about the precise origin of knock-on.  Gross-Pitaevskii simulations \citep{Warszawski:2010individual} point to (1) a vortex proximity effect, which results from an increase in the superfluid velocity at the site of a pinned vortex, when an unpinned vortex passes nearby; and (2) dislodgement by acoustic pulses emitted whenever a vortex repins, which vibrate the superfluid.  Back-of-the-envelope calculations of the magnitude suggest that, for predicted pinning strengths and temperatures in a pulsar, neither mechanism is sufficiently strong to catalyse large-scale unpinning events, (\emph{i.e.} to give $\lambda\tau_{\rm{k}}=1$).  The level of fine-tuning in a model which includes knock-on is quantified.  If knock-on is too strong, most glitches are system-spanning; too weak and single unpinning events dominate, akin to vortex creep \citep{Alpar:1984p6781}.

In closing, we emphasise that the Monte-Carlo simulations described in Sec.~\ref{subsec:avaauto} include all the new physics described in Sec.~\ref{subsec:newphysics}, whereas the master equation (Sec.~\ref{sec:master}) awaits further work.  We cannot explicitly extract the fine-tuning conditions from the automaton.  Hence a solution to Eq.~(\ref{eq:CK}) is worth pursuing.  

The lack of spatial structure in our model is an obvious deficiency.  \cite{CHENG:1988p180,Melatos:2008p204} and many others have cited a richly-connected network of capacitive regions of pinned vortices as the physical origin of broadly distributed glitch sizes.  In this way, even when $\lambda\tau_{\rm{k}}\sim 1$, the entire system of vortices is not `available' to the avalanche, since the knock-on mechanism may peter out once a local `stress reservoir' is exhausted.  We propose two ways to do this:  (1) combine the automaton described in Sec.~\ref{subsec:avaauto} with an automaton [like the one studied in \cite{Warszawski:2008p4510}] that explicitly models nearest-neighbour interactions on a grid of pinning sites; or (2), make the assumption that strong pinning sites are somehow clustered (at grain boundaries in the crust, for example).  We then speculate that the vortex density, and hence the distance of approach of an unpinned vortex to nearby pinned vortices, is generally higher in stronger pinning regions (some evidence of this is described in \cite{Warszawski:2010individual}), so that $\Delta\gamma$ is an increasing function of $E_{\rm{p}}$.  Broadly distributed glitch sizes observed in preliminary simulations that include this dependence (not presented here) are encouraging.  We will pursue this line of thinking in future work.

\bibliographystyle{mn2e}
\bibliography{all_final}

\begin{thebibliography}{75}
\expandafter\ifx\csname natexlab\endcsname\relax\def\natexlab#1{#1}\fi

\bibitem[{Alpar {et~al}\mbox{.}(1981)Alpar, Anderson, Pines, \&
  Shaham}]{Alpar:1981p18}
Alpar M.~A., Anderson P.~W., Pines D., Shaham J., 1981, PNAS, 78, 5299

\bibitem[{Alpar {et~al}\mbox{.}(1984)Alpar, Pines, Anderson, \&
  Shaham}]{Alpar:1984p6781}
Alpar M.~A., Pines D., Anderson P.~W., Shaham J., 1984, ApJ, 276, 325

\bibitem[{{Anderson} \& {Itoh}(1975)}]{Anderson:1975p84}
{Anderson} P.~W., {Itoh} N., 1975, Nature, 256, 25

\bibitem[{Andersson {et~al}\mbox{.}(2007)Andersson, Sidery, \&
  Comer}]{Andersson:2007p196}
Andersson N., Sidery T., Comer G., 2007, MNRAS

\bibitem[{Avise {et~al}\mbox{.}(1984)Avise, Neigel, \& Arnold}]{Avise:1984}
Avise J., Neigel J., Arnold J., 1984, Journal of Molecular Evolution, 20, 99

\bibitem[{Avogadro {et~al}\mbox{.}(2007)Avogadro, Barranco, Broglia, \&
  Vigezzi}]{Avogadro:2007p51}
Avogadro P., Barranco F., Broglia R.~A., Vigezzi E., 2007, PRC, 75, 5

\bibitem[{{Bak} \& {Sneppen}(1993)}]{Bak:1993}
{Bak} P., {Sneppen} K., 1993, PRL, 71, 4083

\bibitem[{Biane {et~al}\mbox{.}(1995)Biane, Durrett, \& Durrett}]{Durrett:1995}
Biane P., Durrett R., Durrett R., 1995, in Lecture Notes in Mathematics, Vol.
  1608, Lectures on Probability Theory, Springer Berlin / Heidelberg, pp.
  97--201

\bibitem[{Blasio \& Lazzari(1998)}]{DeBlasio:1998p127}
Blasio F.~D., Lazzari G., 1998, Nuclear Physics, 633, 391

\bibitem[{{Boerlijst} \& {Hogeweg}(1991)}]{Boerlijst:1991}
{Boerlijst} M.~C., {Hogeweg} P., 1991, Physica D: Nonlinear Phenomena, 48, 17

\bibitem[{{Bouchaud} {et~al}\mbox{.}(1995){Bouchaud}, {Comtet}, \&
  {Monthus}}]{Bouchaud:1995}
{Bouchaud} J., {Comtet} A., {Monthus} C., 1995, Journal de Physique I, 5, 1521

\bibitem[{Carreras {et~al}\mbox{.}(2002)Carreras, Lynch, Dobson, \&
  Newman}]{Carreras:2002p8220}
Carreras B.~A., Lynch V.~E., Dobson I., Newman D.~E., 2002, Chaos, 12, 985

\bibitem[{Carreras {et~al}\mbox{.}(2009)Carreras, Newman, Dobson, \&
  Zeidenberg}]{Carreras:2009p8216}
Carreras B.~A., Newman D.~E., Dobson I., Zeidenberg M., 2009, Chaos, 19, 3107

\bibitem[{Cheng {et~al}\mbox{.}(1988)Cheng, Pines, Alpar, \&
  Shaham}]{CHENG:1988p180}
Cheng K., Pines D., Alpar M., Shaham J., 1988, ApJ, 330, 835

\bibitem[{Chevalier(1993)}]{Chevalier:1993p7949}
Chevalier E., 1993, ApJ, 414, L113

\bibitem[{Consul \& Jain(1973)}]{Consul:1973}
Consul P.~C., Jain G.~C., 1973, Technometrics, 15, pp. 791

\bibitem[{{Cornforth} {et~al}\mbox{.}(2005){Cornforth}, {Green}, \&
  {Newth}}]{Cornforth:2005}
{Cornforth} D., {Green} D.~G., {Newth} D., 2005, Physica D Nonlinear Phenomena,
  204, 70

\bibitem[{Daly \& Porporato(2007)}]{Daly:2007p8447}
Daly E., Porporato A., 2007, PRE, 75, 11119

\bibitem[{{Daly} \& {Porporato}(2010)}]{Daly:2010}
{Daly} E., {Porporato} A., 2010, PRE, 81, 061133

\bibitem[{Dobson {et~al}\mbox{.}(2007)Dobson, Carreras, Lynch, \&
  Newman}]{Dobson:2007p8221}
Dobson I., Carreras B.~A., Lynch V.~E., Newman D.~E., 2007, Chaos, 17, 6103

\bibitem[{Donati \& Pizzochero(2006)}]{Donati:2006p32}
Donati P., Pizzochero P., 2006, Phys. Lett. B, 640, 74

\bibitem[{Donati \& Pizzochero(2003)}]{Donati:2003p97}
Donati P., Pizzochero P.~M., 2003, PRL, 90, 211101

\bibitem[{Donnelly(1991)}]{Donnelly}
Donnelly R.~J., 1991, Quantized vortices in Helium II. Cambridge University
  Press

\bibitem[{Drossel(1996)}]{Drossel:1996}
Drossel B., 1996, PRL, 76, 936

\bibitem[{Epstein \& Baym(1992)}]{Epstein:1992p2172}
Epstein R.~I., Baym G., 1992, ApJ, 387, 276

\bibitem[{{Espinoza} {et~al}\mbox{.}(2011){Espinoza}, {Lyne}, {Stappers}, \&
  {Kramer}}]{Espinoza:2011}
{Espinoza} C.~M., {Lyne} A.~G., {Stappers} B.~W., {Kramer} M., 2011,
  astro-ph.HE/1102.1743

\bibitem[{Feldmann \& Whitt(1998)}]{Feldmann:1998}
Feldmann A., Whitt W., 1998, Performance Evaluation, 31, 245

\bibitem[{Gardiner(2002)}]{Gardiner:2002}
Gardiner C.~W., 2002, {Handbook of stochastic methods : for physics, chemistry
  and the natural sciences}, Springer series in synergetics, 13. Springer

\bibitem[{Glampedakis {et~al}\mbox{.}(2008)Glampedakis, Andersson, \&
  Jones}]{Glampedakis:2008p7}
Glampedakis K., Andersson N., Jones D., 2008, PRL

\bibitem[{{Goss} {et~al}\mbox{.}(1989){Goss}, {Aron}, {Deneubourg}, \&
  {Pasteels}}]{Goss:1989}
{Goss} S., {Aron} S., {Deneubourg} J.~L., {Pasteels} J.~M., 1989,
  Naturwissenschaften, 76, 579

\bibitem[{Hakonen {et~al}\mbox{.}(1998)Hakonen, Avenel, \&
  Varoquaux}]{Hakonen:1998p10787}
Hakonen P., Avenel O., Varoquaux E., 1998, PRL, 81, 3451

\bibitem[{H\"anggi {et~al}\mbox{.}(1990)H\"anggi, Talkner, \&
  Borkovec}]{Hanggi:1990}
H\"anggi P., Talkner P., Borkovec M., 1990, Reviews of Modern Physics, 62, 251

\bibitem[{Harris(1989)}]{Harris}
Harris T.~E., 1989, {The Theory of Branching Processes}. Dover, New York

\bibitem[{{Haskell} {et~al}\mbox{.}(2012){Haskell}, {Pizzochero}, \&
  {Sidery}}]{Haskell:2011}
{Haskell} B., {Pizzochero} P.~M., {Sidery} T., 2012, MNRAS, 420, 658

\bibitem[{{Head}(2000)}]{Head:2000}
{Head} D., 2000, European Physical Journal B, 17, 289

\bibitem[{Huberman \& Glance(1993)}]{Huberman:1993}
Huberman B., Glance N., 1993, PNAS, 90, 7716

\bibitem[{Jagers(2005)}]{Jagers:2005}
Jagers P., 2005, {Branching Processes}. Wiley Online Library

\bibitem[{Jones(1991{\natexlab{a}})}]{Jones:1991p92}
Jones P., 1991{\natexlab{a}}, ApJ, 373, 208

\bibitem[{Jones(1991{\natexlab{b}})}]{Jones:1991p3949}
Jones P.~B., 1991{\natexlab{b}}, ApJ, 373, 208

\bibitem[{{Jones}(1998)}]{Jones:1998p34}
{Jones} P.~B., 1998, PRL, 81, 4560

\bibitem[{Link(2009)}]{Link:2009p9063}
Link B., 2009, PRL, 102, 131101

\bibitem[{{Link}(2012{\natexlab{a}})}]{Link:2012a}
{Link} B., 2012{\natexlab{a}}, MNRAS, 421, 2682

\bibitem[{{Link}(2012{\natexlab{b}})}]{Link:2012b}
{Link} B., 2012{\natexlab{b}}, MNRAS, 422, 1640

\bibitem[{Link {et~al}\mbox{.}(1993)Link, Epstein, \& Baym}]{Link:1993p47}
Link B., Epstein R., Baym G., 1993, ApJ, 402, 285

\bibitem[{Melatos \& Peralta(2007)}]{Melatos:2007p158}
Melatos A., Peralta C., 2007, ApJL, 662, L99

\bibitem[{{Melatos} \& {Peralta}(2010)}]{Melatos:2010turb}
{Melatos} A., {Peralta} C., 2010, ApJ, 709, 77

\bibitem[{Melatos {et~al}\mbox{.}(2008)Melatos, Peralta, \&
  Wyithe}]{Melatos:2008p204}
Melatos A., Peralta C., Wyithe J. S.~B., 2008, ApJ, 672, 1103

\bibitem[{Melatos \& Warszawski(2009)}]{Melatos:2009p4511}
Melatos A., Warszawski L., 2009, ApJ, 700, 1524

\bibitem[{{Mendell}(1991)}]{Mendell:1991}
{Mendell} G., 1991, ApJ, 380, 530

\bibitem[{Mongiov\`\i{} {et~al}\mbox{.}(2007)Mongiov\`\i{}, Jou, \&
  Sciacca}]{Mongiovi:2007p214514}
Mongiov\`\i{} M.~S., Jou D., Sciacca M., 2007, PRB, 75, 214514

\bibitem[{Monthus \& Bouchaud(1996)}]{Monthus:1996p10195}
Monthus C., Bouchaud J., 1996, Journal of Physics A: Mathematical and General,
  29, 3847

\bibitem[{Newman \& Sneppen(1996)}]{Newman:1996p1484}
Newman M., Sneppen K., 1996, PRE, 54

\bibitem[{Peralta {et~al}\mbox{.}(2005)Peralta, Melatos, Giacobello, \&
  Ooi}]{Peralta05}
Peralta C., Melatos A., Giacobello M., Ooi A., 2005, ApJ, 635, 1224

\bibitem[{Peralta {et~al}\mbox{.}(2006)Peralta, Melatos, Giacobello, \&
  Ooi}]{Peralta06a}
Peralta C., Melatos A., Giacobello M., Ooi A., 2006, ApJ, 651, 1079

\bibitem[{{Pizzochero}(2008)}]{Pizzochero:2007p144}
{Pizzochero} P.~M., 2008, in Exotic States of Nuclear Matter, {U.~Lombardo,
  M.~Baldo, F.~Burgio, \& H.-J.~Schulze}, ed., pp. 388--395

\bibitem[{{Pizzochero}(2011)}]{Pizzochero:2011}
{Pizzochero} P.~M., 2011, arxiv:astro-ph.HE/1105.0156

\bibitem[{Pizzochero {et~al}\mbox{.}(1997)Pizzochero, Viverit, \&
  Broglia}]{Pizzochero:1997p45}
Pizzochero P.~M., Viverit L., Broglia R.~A., 1997, PRL, 79, 3347

\bibitem[{Porporato {et~al}\mbox{.}(2004)Porporato, Daly, \&
  Rodriguez-Iturbe}]{Porporato:2004p8491}
Porporato A., Daly E., Rodriguez-Iturbe I., 2004, Am Nat

\bibitem[{{Rinn} {et~al}\mbox{.}(2001){Rinn}, {Maass}, \&
  {Bouchaud}}]{Rinn:2001}
{Rinn} B., {Maass} P., {Bouchaud} J., 2001, PRB, 64, 104417

\bibitem[{Schonfisch \& de~Roos(1999)}]{Schonfisch:1999}
Schonfisch B., de~Roos A., 1999, Biosystems, 51, 123

\bibitem[{{Suki} {et~al}\mbox{.}(1994){Suki}, {Barab{\'a}si}, {Hantos},
  {Pet{\'a}k}, \& {Stanley}}]{Suki:1994}
{Suki} B., {Barab{\'a}si} A., {Hantos} Z., {Pet{\'a}k} F., {Stanley} H.~E.,
  1994, Nature, 368, 615

\bibitem[{Tobita \& Hamielec(1989)}]{Tobita:1989}
Tobita H., Hamielec A., 1989, Macromolecules, 22, 3098

\bibitem[{Tsubota {et~al}\mbox{.}(2000)Tsubota, Araki, \&
  Nemirovskii}]{Tsubota:2000}
Tsubota M., Araki T., Nemirovskii S.~K., 2000, PRB, 62, 11751

\bibitem[{{Tsubota} {et~al}\mbox{.}(2012){Tsubota}, {Kobayashi}, \&
  {Takeuchi}}]{Tsubota:2012}
{Tsubota} M., {Kobayashi} M., {Takeuchi} H., 2012, ArXiv e-prints

\bibitem[{Turcotte(1999)}]{Turcotte:1999}
Turcotte D.~L., 1999, Physics of The Earth and Planetary Interiors, 111, 275

\bibitem[{{van Eysden} \& {Melatos}(2010)}]{VanEysden:2010}
{van Eysden} C.~A., {Melatos} A., 2010, MNRAS, 1459

\bibitem[{Warszawski {et~al}\mbox{.}(2009)Warszawski, Geil, \&
  Wyithe}]{Warszawski:2009p3302}
Warszawski L., Geil P.~M., Wyithe J. S.~B., 2009, MNRAS, 396, 1106

\bibitem[{Warszawski \& Melatos(2008)}]{Warszawski:2008p4510}
Warszawski L., Melatos A., 2008, MNRAS, 390, 175

\bibitem[{Warszawski \& Melatos(2011)}]{Warszawski:2010pulsar}
Warszawski L., Melatos A., 2011, MNRAS, 415, 1611

\bibitem[{{Warszawski} {et~al}\mbox{.}(2012){Warszawski}, {Melatos}, \&
  {Berloff}}]{Warszawski:2010individual}
{Warszawski} L., {Melatos} A., {Berloff} N.~G., 2012, PRB, 85, 104503

\bibitem[{Wheatland(2008)}]{Wheatland:2008p8329}
Wheatland M.~S., 2008, ApJ, 679, 1621

\bibitem[{Wheatland(2009)}]{Wheatland:2009p8258}
Wheatland M.~S., 2009, Solar Physics, 255, 211

\bibitem[{Wong {et~al}\mbox{.}(2001)Wong, Backer, \& Lyne}]{Wong:2001p192}
Wong T., Backer D., Lyne A., 2001, ApJ, 548, 447

\bibitem[{{Worrell} {et~al}\mbox{.}(2002){Worrell}, {Cranstoun}, {Echauz}, \&
  {Litt}}]{Worrell:2002}
{Worrell} G., {Cranstoun} D., {Echauz} J., {Litt} B., 2002, NeuroReport, 13

\bibitem[{Wu {et~al}\mbox{.}(2010)Wu, Zhou, Xiao, Kurths, \&
  Schellnhuber}]{Wu:2010}
Wu Y., Zhou C., Xiao J., Kurths J.~A., Schellnhuber H.~J., 2010, Proceedings of
  the National Academy of Sciences, 107, 18803

\end{thebibliography}

\appendix
\section{Superfluid angular momentum and vortex motion}\label{app:dL}
Consider the angular momentum of a superfluid, projected onto the rotation axis, $L_z$, containing a smooth, continuous distribution of vortices,
\begin{equation}
\label{eq:Lz}
L_z = 2\pi R\rho\int_0^R dr\,r |\mathbf{r}\times \mathbf{v}_{\rm{s}}~|,
\end{equation}
where $\rho$ is the superfluid density, $\mathbf{v}_{\rm{s}}$ is the superfluid velocity, and we assume that the system is uniform in the direction of the rotation axis, and circularly symmetric in the transverse plane.  We can rewrite $|\mathbf{v}_{\rm{s}}|$ in terms of the number of vortices $N(r)$ at radii up to $r$, 
\begin{equation}
|\mathbf{v}_{\rm{s}}| =\frac{\kappa N(r)}{2\pi r}~,
\end{equation}
such that Eq.~(\ref{eq:Lz}) becomes
\begin{equation}
L_z = 2\pi R\rho\int_0^R dr\frac{r\kappa N(r)}{2\pi}~.
\end{equation}
Assume now that $\Delta N$ vortices move from radius $R_1$ to $R_2$, with $R_2-R_1=\alpha R$, while all other vortices remain stationary.  $L_z$ then becomes
\begin{equation}
L_z-\Delta L_z = R\kappa\rho\left\lbrace\int_0^{R_1} dr\,rN(r)+\int_{R_1}^{R_2}dr\,r\left[N(r)-\Delta N\right]+\int_{R_2}^{R}dr\,rN(r)\right\rbrace~.
\end{equation}
For $R_1\approx R$, we get
\begin{equation}
\Delta L_z \approx -\rho\kappa\alpha \Delta N R^3~.
\end{equation}

\section{Relaxation to equilibrium}\label{sec:knockappa}
In this appendix, we solve (\ref{eq:masterg}) explicitly for
$g(E_{\rm{p}},t)$ given $\phi(E_{\rm{p}})$ as in Eq.~(\ref{eq:phiofE}), taking
$\Gamma_0$ and $E_{\rm{p}}=E_0$ to be constants.
This latter assumption reflects the observational fact that
the glitch size and waiting-time distributions appear to
converge to a quasistationary state over years,
i.e.\ over an interval much shorter than the 
spin-down time-scale $\Omega/\dot{\Omega}$
\citep{Melatos:2008p204}.
The reason for solving analytically is to find explicitly
the time-scales over which relaxation to equilibrium occurs. 
We Laplace transform Eq.~(\ref{eq:masterg}), defining 
$g(E_{\rm{p}},s) = \int_0^\infty dt'\, e^{-st'} g(E_{\rm{p}},t')$,
and rearrange to obtain
\begin{equation}
 g(E_{\rm{p}},s) = \frac{g(E_{\rm{p}},t=0)}{s+\Gamma_0 e^{-\beta E_{\rm{p}}}}+\frac{\Gamma_0 \omega(s) \phi(E_{\rm{p}})}{s+\Gamma_0 e^{-\beta E_{\rm{p}}}}~.
\label{eq:knockappa1}
\end{equation}
Integrating (\ref{eq:knockappa1}) over all $E_{\rm{p}}$,
and noting $s^{-1} = \int_0^\infty dE' \, g(E',s)$,
we find
\begin{equation}
 \Gamma_0 \omega(s)
 =\left[\frac{1}{s}-\int_0^\infty dE' \, \frac{g(E',t=0)}{s+\Gamma_0 e^{-\beta E_{\rm{p}}'}} \right]  \left/ 
  \int_0^\infty dE' \,
  \frac{\phi(E')}{s+\Gamma_0 e^{-\beta E_{\rm{p}}'}}~.
 \right.
\label{eq:knockappa2}
\end{equation}
Equation (\ref{eq:knockappa2}) is general. Let us evaluate it
for the reasonable special case $g(E,t=0) = \phi(E)$,
with $\phi(E)$ given by the top-hat function (\ref{eq:phiofE}).
One then finds
\begin{equation}
 \Gamma_0\omega(s)
 =
 \frac{2\beta E_0}
  {\ln\left( \frac{se^{2\beta E_0}+\Gamma_0}{s+\Gamma_0} \right)}
 - 1~,
\label{eq:knockappa3}
\end{equation}
and hence
\begin{equation}
 g(E_{\rm{p}},s)
 =
 \frac{\beta H(E_{\rm{p}}) H(2E_0 - E_{\rm{p}})}
  { (s+\Gamma_0 e^{-\beta E_{\rm{p}}})  
    \ln\left( \frac{se^{2\beta E_0}+\Gamma_0}{s+\Gamma_0} \right) }~.
\label{eq:knockappa4}
\end{equation}
We recover the initial distribution (\ref{eq:phiofE})
as expected from the initial value theorem
$g(E,t=0^+) = \lim_{s\rightarrow\infty} sg(E,s)$.
We also recover the stationary distribution (\ref{eq:knock8})
from the final value theorem, which says that 
$g(E,t\rightarrow\infty) = \lim_{s\rightarrow 0} sg(E,s)$
because all the poles of $sg(E,s)$ lie in the left half plane.
We can invert the Laplace transform (\ref{eq:knockappa4}) 
by evaluating the Bromwich integral
\begin{equation}
 g(E_{\rm{p}},t)
 =
 \frac{1}{2\pi i} 
 \int_{\gamma-i\infty}^{\gamma+i\infty}
 ds\, e^{st} g(E_{\rm{p}},s)~,
\label{eq:knockappa5}
\end{equation}
where $s$ is now a complex number and the integral is along the
line $Re(s)=\gamma$, where $\gamma$ is taken to the right of
all the poles of $e^{st} g(E,s)$ in the complex plane.
\end{document}